\documentclass[ aps, amsmath,amssymb,preprint,altaffilletter,superscriptaddress,floatfix
]{revtex4-1}

\usepackage{amsmath}          
\usepackage{amsfonts}          
\usepackage{amssymb}           
\usepackage{graphicx}
\usepackage{subfigure}          
\usepackage{multirow}			
\usepackage[table]{xcolor}		
\usepackage{textcomp}			

\begin{document} 
\title{Role of ferroelectric polarization during growth of highly strained ferroelectrics revealed by in-situ x-ray diffraction}
\author{Rui Liu}
\affiliation{Department of Physics and Astronomy, Stony Brook University, \\Stony Brook, NY 11794-3800 USA.}
\author{Jeffrey G. Ulbrandt}
\affiliation{Department of Physics and Materials Science Program, University of Vermont, \\Burlington, Vermont 05405, USA.}
\author{Hsiang-Chun Hsing}
\author{Anna Gura}
\author{Benjamin Bein}
\author{Alec Sun}
\author{Charles Pan}
\author{Giulia Bertino}
\author{Amanda Lai}
\author{Kaize Cheng}
\author{Eli Doyle}
\affiliation{Department of Physics and Astronomy, Stony Brook University, \\Stony Brook, NY 11794-3800 USA.}
\author{Kenneth Evans-Lutterodt}
\affiliation{Brookhaven National Laboratory, Upton, NY 11973 USA}
\author{Randall L. Headrick}
\affiliation{Department of Physics and Materials Science Program, University of Vermont, \\Burlington, Vermont 05405, USA.}
\author{Matthew  Dawber}
\email{matthew.dawber@stonybrook.edu}
\affiliation{Department of Physics and Astronomy, Stony Brook University, \\Stony Brook, NY 11794-3800 USA.}

\begin{abstract}
Strain engineering of perovskite oxide thin films has proven to be an extremely powerful method for enhancing and inducing ferroelectric behavior. In ferroelectric thin films and superlattices, the polarization is intricately linked to crystal structure, but we show here that it can also play an important role in the growth process, influencing growth rates, relaxation mechanisms, electrical properties and domain structures. We have studied this effect in detail by focusing on the properties of BaTiO$_{3}$ thin films grown on very thin layers of PbTiO$_{3}$ using a combination of x-ray diffraction, piezoforce microscopy, electrical characterization and rapid in-situ x-ray diffraction reciprocal space maps during the growth using synchrotron radiation. Using a simple model we show that the changes in growth are driven by the energy cost for the top material to sustain the polarization imposed upon it by the underlying layer, and these effects may be expected to occur in other multilayer systems where polarization is present during growth. Our research motivates the concept of polarization engineering during the growth process as a new and complementary approach to strain engineering.
\end{abstract}
 
\maketitle

Strain engineering of perovskite oxide thin films has proven to be an extremely powerful approach for enhancing and inducing ferroelectric behavior. Some of the most notable achievements in recent years have included the enhancement of polarization in strained BaTiO$_{3}$ by 270\%\cite{Choi04}, discovery of a super- tetragonal phase of BiFeO$_{3}$ \cite{Bea09} and a strain enabled electric control over magnetization in EuTiO$_{3}$\cite{Lee10,Ryan13}. In ferroelectric thin films and multilayers, the polarization is intricately linked to crystal structure,  so that strain and electrostatic boundary conditions have considerable impact on the magnitude of the polarization and the arrangement of polarization domains \cite{Junquera03,Dawber03,Lichtensteiger05,Streiffer02,Fong04,Fong06,Dawber05,Dawber07,McQuaid11, Zubko12,Lichtensteiger14,Bein15,Lichtensteiger16,Yadav16,Hong17,Damoradan17,Hadjimichael18}. In certain strained ferroelectrics, for example BaTiO$_{3}$ (BTO) or PbTiO$_{3}$ (PTO) grown epitaxially on SrTiO$_{3}$ (STO), the ferroelectric transition temperature can lie above the growth temperature of the film \cite{Sinsheimer13,Bein15}, raising the question: Is it possible to engineer material properties of ferroelectric thin films during growth, not only by strain, but also polarization? Here we demonstrate that the answer to this question is an emphatic ``yes'' and provide insight into the mechanism through which this occurs by performing in-situ x-ray diffraction during growth at the National Synchrotron Light Source-II (NSLS-II).

\begin{figure}[ht!]
	\begin{center}
		\includegraphics[width=8.5cm]{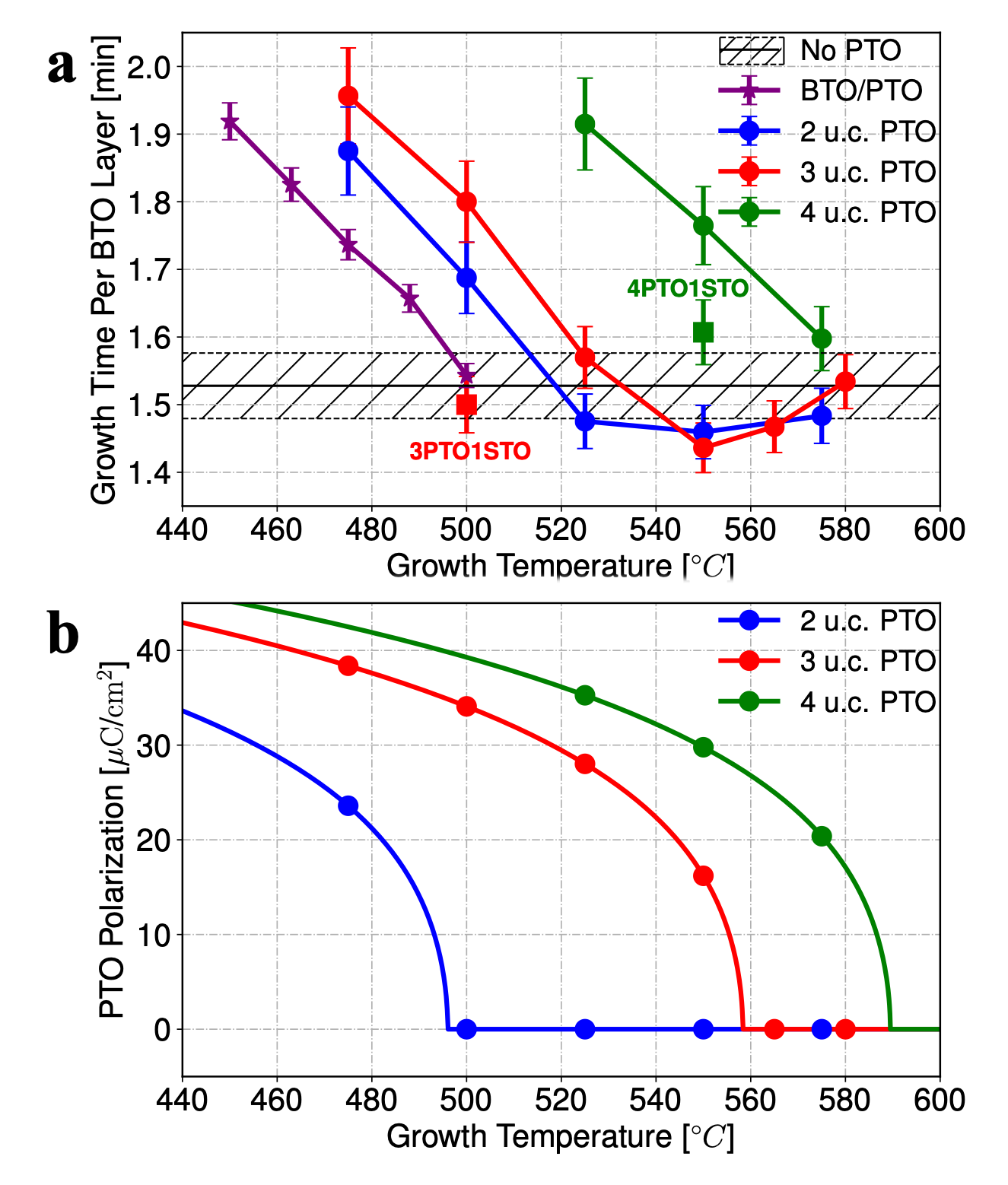}
		
		\caption[Growth rates of BTO and polarization of PTO changes vs temperature]{Growth rates and polarization associated with phase transition: 
			(a) Growth time per BTO layer plotted against growth temperatures in 20 nm pure BTO films (black), 50 nm BTO/PTO superlattices (purple), 20 nm BTO films grown on ultrathin PTO films (circle data point markers, blue: 2 u.c. layers PTO, red: 3 u.c. layers PTO, green: 4 u.c. layers PTO), and  BTO films with an additional 1 u.c. layer STO on top of PTO ultrathin films (square data point markers, red: 3 u.c. layers PTO, green: 4 u.c. layers PTO).  (b) Polarization of ultrathin PTO films under growth conditions calculated according to Equation (1). Source data are provided as a Source Data file.}
		\label{rate_img}
	\end{center}
	\vspace{-20pt}
\end{figure}

A dramatic demonstration of the effect of ferroelectric polarization during growth can be seen in BaTiO$_{3}$/PbTiO$_{3}$ (BTO/PTO) superlattices on SrTiO$_{3}$ substrates. Due to the high compressive strain, particularly that imposed upon  BTO, both BTO and PTO will have highly elevated ferroelectric transition temperatures, though this effect does compete with supression of their ferroelectriciy due to depolarization field when they are ultrathin. Thus for practical deposition temperatures, both of these materials will start out paraelectric and become ferroelectric at a thickness defined by the growth temperature as they are grown. Here we focus on a series of 3 unit cell (u.c.)/3 unit cell BTO/PTO superlattices that were grown in the temperature range from  450$^{\circ}$ C to 500$^{\circ}$ C.  The growth rates of the BTO layers were found to change dramatically over this temperature range (purple data, Fig. \ref{rate_img} (a)). In general growth rates are mostly determined by the incident flux of material and are thus not expected to vary greatly with temperature and over many years of growing superlattices containing PTO \cite{Dawber05,Dawber07,Callori12,Sinsheimer12} this is the first time we have observed significant temperature dependence of the growth rate of one of the layers. In particular we do not see significant temperature dependence of the growth rate when dielectric layers, such as SrTiO$_{3}$ or metallic layers such as SrRuO$_{3}$ are used. Further, when the growth rate of BTO thin films ($\approx$ 20nm thickness) at different temperatures (black data, Fig. \ref{rate_img} (a)) is measured it does not appear to depend on temperature. We thus suppose that the existence of the 3 u.c. layers of PTO plays a key role in determining the BTO growth rate within the superlattice. Previous studies of ultrathin PTO films found that the ferroelectric phase transition temperature of 3 u.c. films of PTO is close to our growth temperature\cite{Fong06}, so a plausible hypothesis is that the change in growth rate of BTO may be associated with the large changes of polarization with temperature that PTO should display in the vicinity of the ferroelectric phase transition.

To verify the hypothesis, ultrathin PTO films were grown at temperatures in the vicinity of the ferroelectric phase transition temperature and then BTO films were grown on top of them to a thickness of 20nm. In other words, the BTO films were grown on ``substrates'' with various ferroelectric polarization values. A 20nm thick SrRuO$_{3}$ (SRO) electrode was grown beneath all of the films to enable electrical measurements. The surface quality of the grown films was checked using atomic force microscopy (AFM) (supplementary Fig. 2). Crystal structures and growth rates were determined by X-Ray diffraction $\theta-2\theta$ scans and low angle reflectivity scans on a Bruker D8-Discover high-resolution X-Ray Diffractometer (supplementary Fig. 1). Three series of samples were prepared with 2, 3 and 4 u.c. layer thick PTO films, and the measured average growth rates for these films as function of temperature in between 450\textdegree C and 580\textdegree C are shown in (Fig. \ref{rate_img} (a) blue, red and green). We did not examine growth above 600$^{\circ}$ as there is substantial Pb loss from the PTO at these higher growth temperatures. The growth rate displayed is the average growth rate for the growth of the entire 20nm film and it can be seen that overall the growth rates are somewhat slower for the films than the 3 u.c. grown in the BTO/PTO superlattices. This change in growth rate as the film thickness increases is confirmed by our in-situ x-ray diffraction measurements (Supplementary Figure 6)

To understand the link between the growth temperature and PTO thickness and the ferroelectric polarization of PTO at the growth condition, the Landau approach of Pertsev et al\cite{Pertsev98} was used. In this approach the out of plane polarization of the ultra thin strained PTO can be calculated from minimizing

\begin{equation}
G=a_{3}^*P^{2}+a_{33}^*P^{4}+a_{111}P^{6}+\frac{u_{m}^{2}}{s_{11}+s_{12}}+\frac{\lambda_{eff}}{d\epsilon_{0}}P^{2}
\end{equation}

where $a_{3}^{*}=a_{1}-u_{m}\frac{2Q_{12}}{s_{11}+s_{12}}$ and $a_{33}^*=a_{11}+\frac{Q_{12}^{2}}{s_{11}+s_{12}}$. 

\begin{figure}[ht!]
	\begin{center}
		\includegraphics[width=14cm]{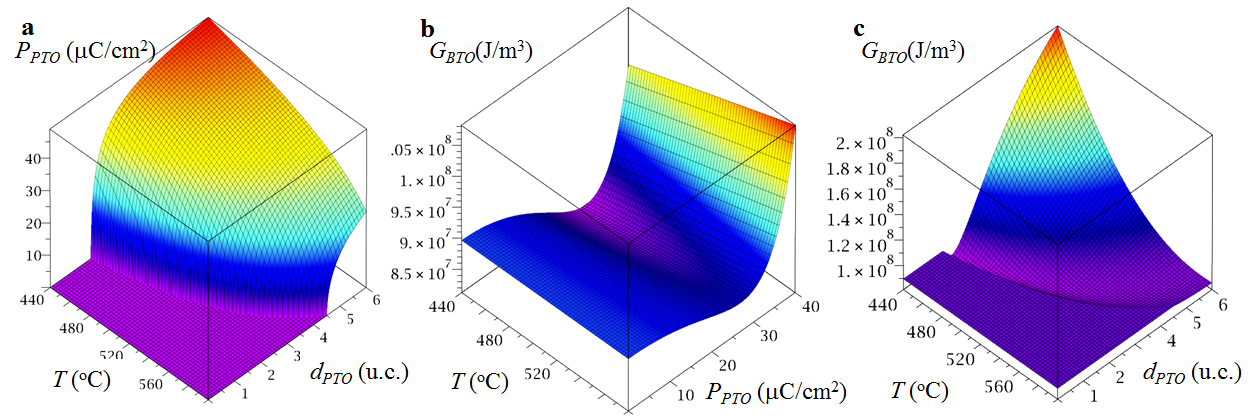}
		
		\caption[Explanation of growth rate through free energy of BTO]{Explanation of growth rate through free energy of BTO: (a) Polarization of PTO as a function of temperature and PTO thickness $d$, (b) Free energy of BTO as a function of temperature and polarization, which in our model is assumed to be the polarization of the underlying PTO, (c) Free energy of BTO as a function of temperature and PTO thickness $d$.}
		\label{BTOenergy}
	\end{center}
	
\end{figure}

We have used the effective screening length $\lambda_{eff}$ as an adjustable parameter to approximately match the transition temperature observed in experiment; the value we used for all calculations shown in the paper was $5\times10^{-13}$m. The predicted polarization of the PTO film under growth conditions is shown in Fig. \ref{rate_img} (b) and Fig. \ref{BTOenergy} (a).

The temperature dependence of growth time per BTO layer in each series (Fig. \ref{rate_img} (a)) shows a strong correlation with the expected polarization of PTO during the growth (Fig. \ref{rate_img} (b)), suggesting that the growth rate of the BTO film changes with the ferroelectric polarization of PTO film on which it grows. If one considers the point at which growth rates begin to depend on temperature as the ferroelectric transition temperature we see that the transition temperature increases as the film becomes thicker, which has been confirmed previously\cite{Streiffer02}. The growth rate of the BTO film does not change when the PTO film under it is in the paraelectric state during growth, while it takes longer to grow when the ferroelectric polarization of PTO film increases. A simple way for rationalizing this result is to consider the additional energy required to change the thickness of the film. Assuming all other things are equal, the additional polarization free energy to add a unit cell of BTO is directly proportional to $G_{BTO}$ (Fig. \ref{BTOenergy} (b)). The initial boundary condition we impose on BTO growing on PTO is that the polarization is continuous across the interface. We therefore take the calculated polarization for the thin PTO layer $P_{PTO}$ (Fig. \ref{BTOenergy} (a)) as the polarization of BTO and calculate $G_{BTO}$ as a function of temperature and the thickness of PTO, $d$ (Fig. \ref{BTOenergy} (c)) using again the approach of Pertsev et al.\cite{Pertsev98}. While the approximation of continuous polarization will quickly break down as the layer grows requiring additional terms to be added to the expansion \cite{Okatan09}, we use the continuous polarization approach to find the simplest way to understand the observed slow down of the growth rate.

Two considerations are important in elevating the bulk ferroelectric free energy of BTO to be a significant factor in the growth, compared to the normally more relevant surface and interface energies, one is the sharp increase in $G_{BTO}$ as a function of $P$ once it exceeds the optimal value for BTO, and the other is the baseline increase in free energy to around $1\times10^{8}\,\mathrm{J/m^{3}}$ due to the mismatch strain imposed on BTO.

Our calculation reveals that there is actually a small region of parameter space close to the phase transition of PTO  where polarization is fairly small and there is a decrease in energy that we might expect leads to a faster growth rate. However, for larger polarizations at lower temperatures and larger PTO thicknesses the free energy of BTO is considerably higher as the PTO tries to impose a higher polarization on the BTO than ideal, resulting in an additional energy cost to increase the thickness of the film. Careful inspection of Fig. \ref{rate_img} (a), reveals that is precisely the behavior which is experimentally observed.

In principle the slow down in growth rate for large PTO polarization is not reliant on the ferroelectric nature of BTO, a dielectric material should similarly experience an energy penalty when polarization is imposed on it. However, for the frequently studied case of SrTiO$_{3}$ grown on PTO on SrTiO$_{3}$  substrates the energy penalty for the polarization values considered here is an order of magnitude less, due to SrTiO$_{3}$ having a large dielectric constant and the lack of a strain induced increase in the bulk free energy, and thus there is not a significant change in the growth rate of SrTiO$_{3}$ due to imposed polarization, which is what we have observed in experiment. However for dielectric layers with lower dielectric constant, or under significant strain we should expect this effect to be observed.

Further evidence for our hypothesis was obtained by growing films in which a single unit cell of STO was grown in between the PTO and the BTO, and these films show a marked reduction in the growth time per layer, as expected by the reduction of polarization that should occur due to the insertion of the STO layer. The connection between BTO growth rate and the ferroelectric polarization of the PTO film underneath also provides an explanation for why the growth rate of BTO film changes with temperature in BTO/PTO superlattices while it remains stable in pure BTO films.

\begin{figure}[ht!]
	\begin{center}
		\includegraphics[width=8.5cm]{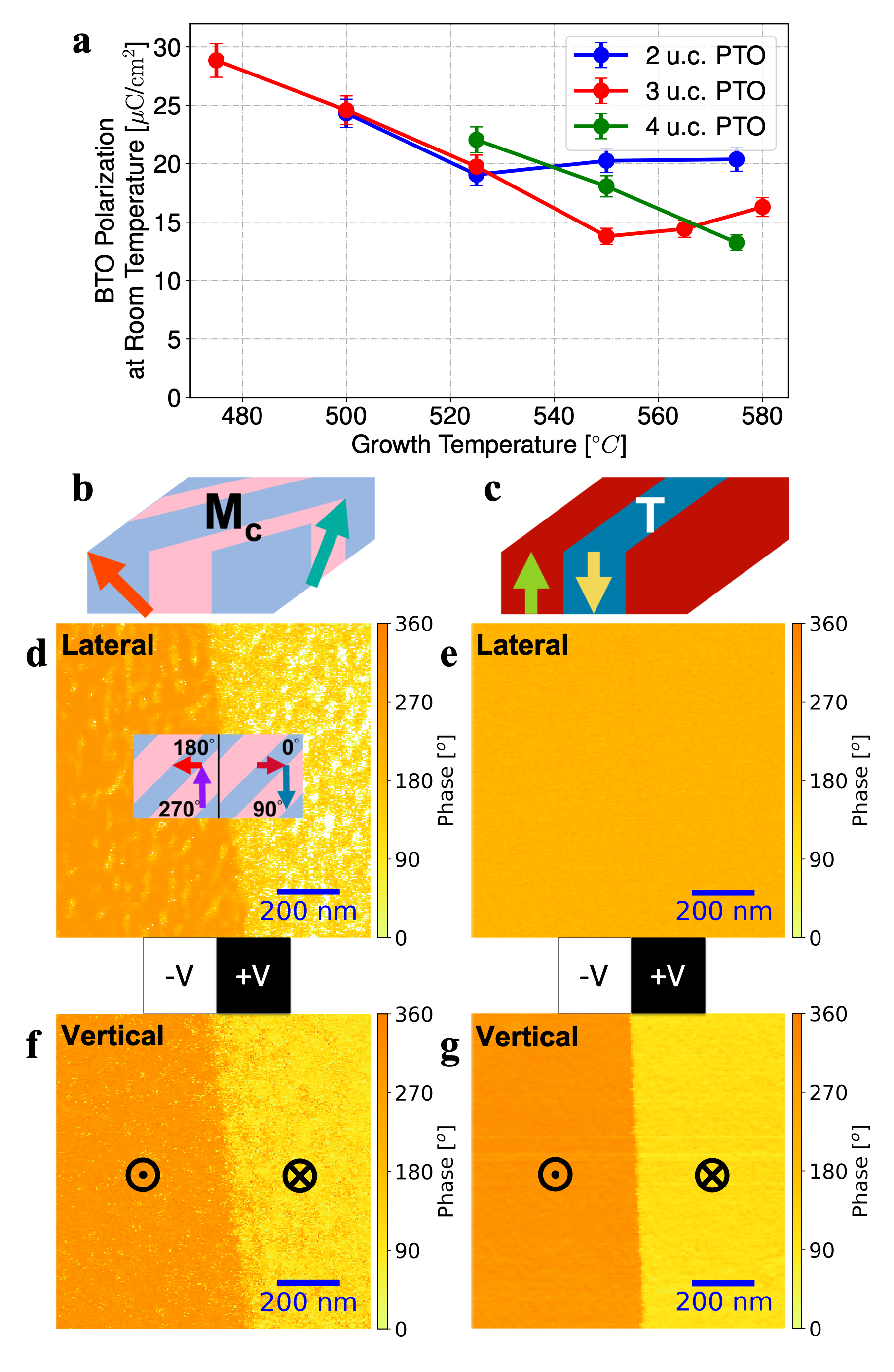}
		
		\caption[lateral and vertical PFM on films grown on ferroelectric/paraelectric PTO and schematic views of the ferroelectric domain structures]{((a) Polarization of BTO films grown on ultrathin PTO films measured at room temperature (blue: 2 u.c. layers PTO, red: 3 u.c. layers PTO, green: 4 u.c. layers PTO). Source data are provided as a Source Data file. (b-g) Ferroelectric domains measured by vector PFM and their possible arrangements: 
			schematics of Monoclinic M$_c$ (b) and Tetragonal (c) domain structures of BTO films grown on paraelectric/ferroelectric 3 u. c. layers PTO films, respectively. Two stable states of uniform and opposite polarization, ``Up'' and ``Down'', were achieved by applying negative voltage in the out of plane direction on the left half side and positive voltage in the out of plane direction on the right half side of each film. The larger micron scale domain structure is that written with the AFM tip. The smaller scale texture in the images is from the naturally occurring domains which are on the 20-60nm length scale. In contrast with vertical response (f, g),  there is no lateral phase PFM response (e) for the film grown on ferroelectric PTO. On the other hand, the lateral cross section of domains (inserted image in (d)) in the film grown on paraelectric PTO indicates two polarization orientations with 90$^\circ$ domain walls in each state, and the lateral phase PFM image (d) does show the 180$^\circ$/270$^\circ$ (``Up'' state) and 0$^\circ$/90$^\circ$ (``Down'' state) domain configuration with four orientations.}
		\label{PFM_img}
	\end{center}
	\vspace{-20pt}
\end{figure}

The post-deposition functional properties of the films were measured at room temperature. One very intriguing observation is that the room temperature out-of-plane polarization of the film as measured by hysteresis measurements also depends on the magnitude of the PTO polarization during growth (Fig. \ref{PFM_img} (a)). Taking films with 3 u.c. layers of PTO as an example (red data, Fig. \ref{PFM_img} (a)), the out-of-plane polarization of  a BTO film grown on ferroelectric PTO simultaneously increases with PTO polarization and can be as large as doubled (29 $\mu\mathrm{C}/\mathrm{cm}^{2}$) compared with the one of BTO film grown on paraelectric PTO film (14 $\mu\mathrm{C}/\mathrm{cm}^{2}$).  Vector piezoforce microscopy (PFM) was then performed in order to determine the piezoelectric response in micro-scale in both lateral and vertical directions. The study of two BTO films grown on 3 u.c layers paraelectric/ferroelectric (grown at 550$^{\circ}$ C/500$^{\circ}$ C) PTO is presented in Fig. \ref{PFM_img}. Two stable states of uniform and opposite polarization, ``Up'' and ``Down'', were achieved by applying negative vertical voltage on the left and positive vertical voltage on the right. In the vertical phase PFM phase response (Fig. \ref{PFM_img} (f,g)), the two films displayed the same behavior; they have a uniform polarization direction in each ``Up'' or ``Down'' state and the two states have 180$^\circ$ difference in domain orientations. However, different behaviors were seen in the lateral PFM phase response. No lateral piezoelectric response (Fig. \ref{PFM_img} (e)) was observed in the films grown on ferroelectric PTO, while complex domain patterns (Fig. \ref{PFM_img} (d)) were seen in the other film grown on paraelectric PTO. In each of the ``Up'' or ``Down'' state, two types of polarization domains with 90$^\circ$ domain walls were observed, and the orientations of polarization in each domain were flipped while the state changed from ``Up'' to ``Down''. The 3D Domain configuration inferred by combining the lateral and vertical PFM response is that the BTO film grown on paraelectric PTO presents low symmetry monoclinic M$_c$ polarization domains, while tetragonal polarization domains were observed in the BTO film grown on ferroelectric PTO. An additional observation made during the PFM measurement is that those films grown on ferroelectric PTO substrates have an electromechanical resonance frequency about 10 \% lower than those grown on paraelectric substrates (Supplementary Fig. 5), implying a difference in the elastic properties of the samples, which in turn have an influence on the electromechanical resonance frequency.

\begin{figure*}[ht!]
	\begin{center}
		\includegraphics[width=16cm]{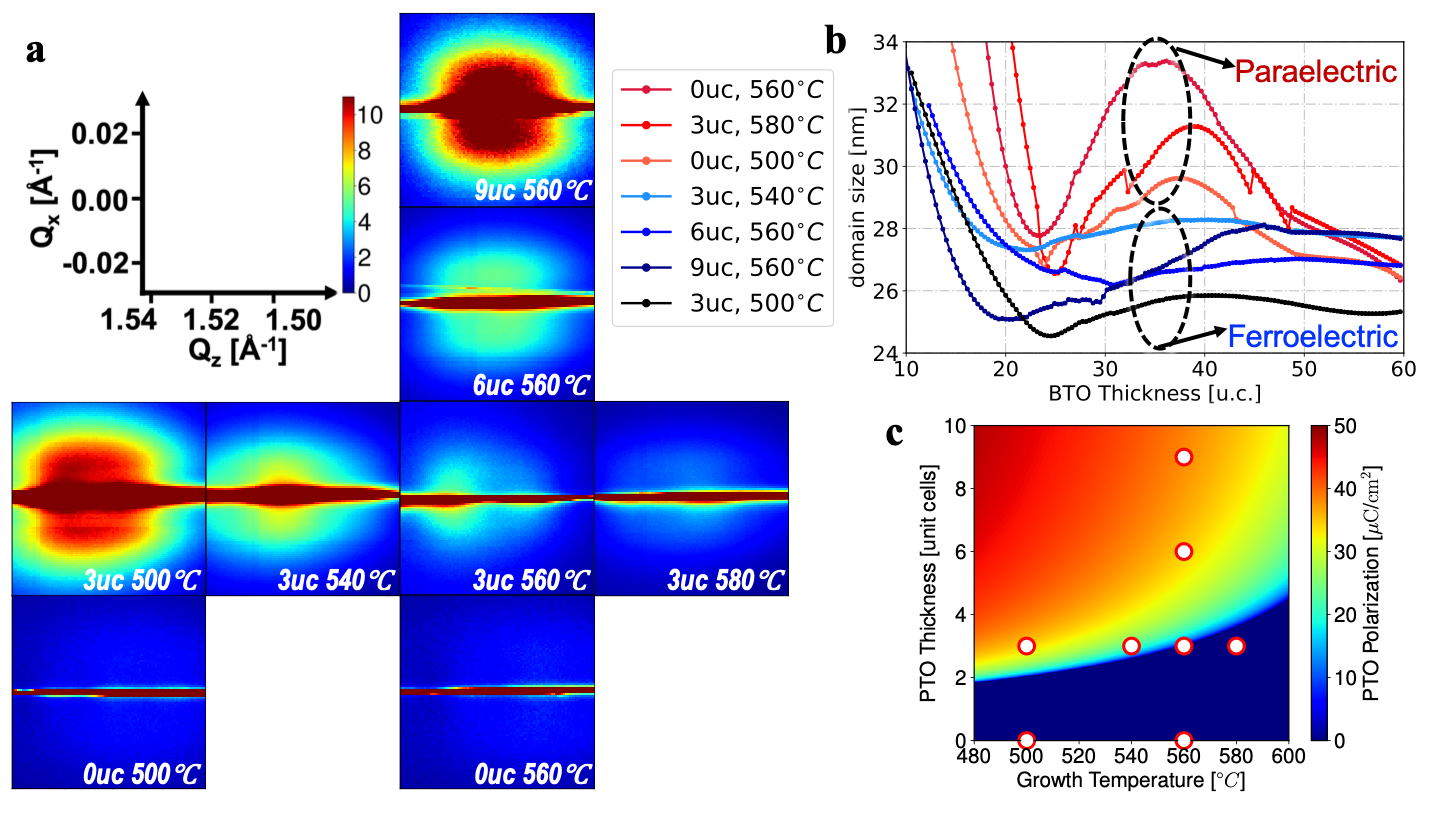}
		
		\caption[In-situ X-ray scans around (0\,0\,1) peak]{In-situ X-ray scans around (0\,0\,1) peak: 
			 (a) Examples of reciprocal space maps obtained by in-situ X-ray near (0\,0\,1) peak after 10nm BTO films were grown. The polarization of PTO ultra-thin films was plotted in (c). The effect of polarization on domains evolution can be seen either in temperature or PTO thickness direction.  (b) Evolution of domain sizes plotted against BTO growth time. The domain sizes of BTO films grown on ferroelectric PTO reach a stable value in the early stage of growth, while the ones grown on paraelectric PTO experience a dramatic evolution. The domain structures of ultrathin PTO films tend to carry over to BTO film if grown on ferroelectric PTO, which makes the domain evolution more moderate. Source data are provided as a Source Data file. (c) Polarization of PTO ultra-thin films calculated via Landau theory plotted against PTO thickness and growth temperature. The red circles corresponded to the samples grown in this experiment.}
		\label{ISR_img}
	\end{center}
	\vspace{-20pt}
\end{figure*}

Due to the  large compressive strain imposed on a BTO film (-2.69 \%) grown on STO substrate, it is no surprise that the strained BTO film starts to relax at some point during the growth, forming some in-plane polarization domains. However our experimental results suggest the ferroelectric polarization of the PTO film underneath can prevent or delay this relaxation and help the BTO film stay in a strained state, protecting the tetragonal polarization domains.

To verify this hypothesis and gain more insight into the mechanisms at play we performed a series of in-situ x-ray diffraction experiments during growth of the model PTO/BTO system at the 4-D beamline at the NSLS-II synchrotron at BNL. These kinds of experiments are a powerful tool for gaining insight in to thin film growth in general\cite{Vlieg88,Levine89,Renaud03,Renaud09,Ozyadin05,Rainville15,Ju19} and  more specifically the evolution of polarization and domain structure in growing ferroelectric films. \cite{Murty02,Fong06,Sinsheimer13,Bein15} Fig. \ref{ISR_img}(c) shows the calculated polarization from Equation (1) for the thin PbTiO$_{3}$ layer $P_{PTO}$  plotted as a function of thickness and temperature, and provides a map to understand the parameters we chose for the PTO layer in our in-situ growth experiments. BTO films of 24nm thickness were grown on ultrathin PTO films with different thickness and different growth temperatures. Both the PTO and BTO were deposited while x-ray diffraction was performed. All films were grown on STO substrates with a 20nm SRO electrode grown ex-situ. Prior to the main set of experiments discussed here the evolution of the anti-Bragg peak (0\,0\,$\frac{1}{2}$) (to minimize bulk Bragg diffraction) \cite{Vlieg88,Murty02,Chinta12}) was measured to calibrate the growth rates for both BTO and PTO films (supplementary Fig. 6). In the following experiments the same scanning technique was used as that of Bein et al.\cite{Bein15} for the study of ferroelectric-dielectric BaTiO$_{3}$/SrTiO$_{3}$ superlattices. The technique allows us to achieve a rapid reciprocal space map in 15s, which is much shorter than the time taken to deposit a single unit cell of material (a few minutes). In each scan, the in-plane angle $\phi$ moves continuously through a given angular range, which corresponds to perform a rocking curve. Thus the measured intensity of each pixel is the integrated intensity over the rocking curve at that pixel. Accordingly one in-plane momentum transfer direction ($Q_y$) is integrated, while diffraction information in the other two directions ($Q_x$ and $Q_z$) can be obtained. Further explanation of the technique can be found in Methods and Supplementary materials.

\begin{figure*}[ht!]
	\begin{center}
		\includegraphics[width=16cm]{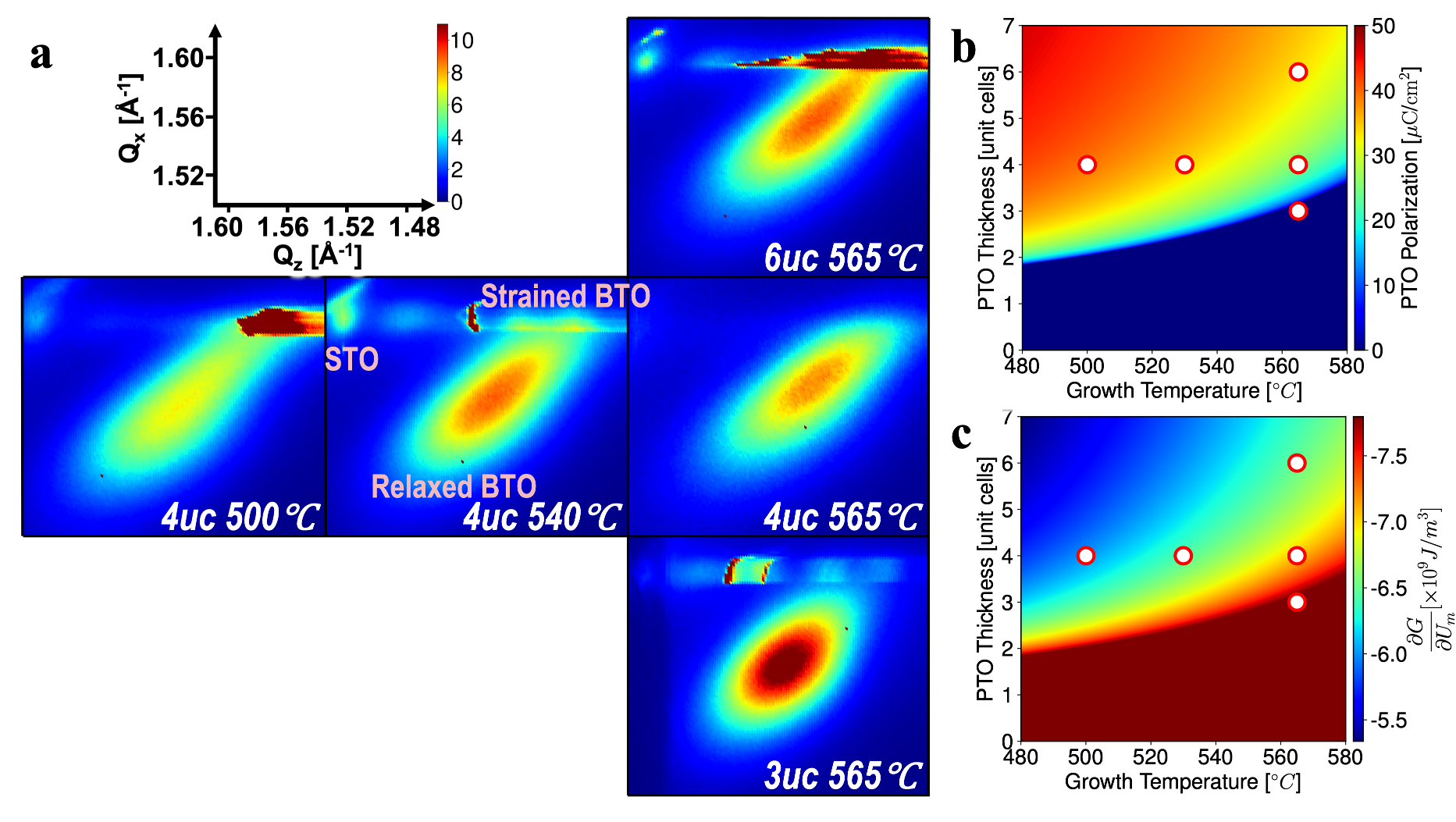}
		
		\caption[Examples of (1\,0\,1) in-situ X-ray scan images]{(a) Examples of reciprocal space maps obtained by in-situ X-ray near (1\,0\,1) peak: 
			Single scan images around (1\,0\,1) peak after 10nm BTO is grown on top of different thickness of PTO grown at different temperature, (b) Polarization  of ultrathin PTO films and (c) $\frac{\partial G}{\partial u_{m}}$  of BTO films calculated via Landau theory plotted against thicknesses and temperatures. The red circles in (b) and (c) correspond to the samples  in (a).}
		\label{RSM_img}
	\end{center}
	\vspace{-20pt}
\end{figure*}

The idea that the ferroelectric polarization of PTO carries over to BTO is a central part of our model. Depending on the electrical boundary conditions underneath them, Fong et al\cite{Fong04,Fong06},  found that ultra-thin PTO layers have a stripe domain structure which produce diffraction features. In our experiment when we performed (0\,0\,1) maps during the growth of BTO on PTO layers with different ferroelectric polarization states we found that the scattering from stripe domains is strongly dependent on the thickness of the PTO layer and the growth temperature. Reciprocal space maps made when 10nm BTO films grown on different PTO ``substrates'' are presented in Fig. \ref{ISR_img}(a). The calculated polarization under growth conditions of PTO ``substrates'' for each film in Fig. \ref{ISR_img}(a) corresponds to the red circles in Fig. \ref{ISR_img}(c). BTO films grown on ``ferroelectric substrates'' show clear and strong domain scatterings, while no obvious domain scattering was observed when the films are grown on ``paraelectric substrates''. Two pure BTO films without PTO layers (labelled as ``0 uc'' were grown at different temperatures. The similarity of the domain scattering from these samples with the ones grown on ``paraelectric substrates'' suggest again that this phenomenon is caused by ferroelectric polarization rather than purely temperature. The domain size evolution during the growth was plotted in Fig. \ref{ISR_img}(b). The domain sizes of BTO films grown on ``ferroelectric substrates'' stabilized faster since the ferroelectric domains of PTO are carried over to BTO,  while the ones grown on ``paraelectric substrates'' go though a more dramatic evolution.

To study strain relaxation we carried out another set of maps, this time around the (1\,0\,1) Bragg peak. Fig \ref{RSM_img}  (a) shows close-ups of the relaxed feature on a number of samples. These figures have been arranged according to their position on a plot of thickness vs growth temperature as shown in Fig. \ref{RSM_img} (b) and (c).  As in Fig. \ref{ISR_img}, the polarization of PTO ``substrates'' of each films in (a) corresponds to the red circles in (b). The  thickness oscillations in these images correspond to the strained part of the BTO film which is constrained in plane to the STO substrate. In addition to this we observe a diffraction feature associated with relaxed BTO. The relaxed parts of the BTO films grown on ferroelectric PTO is connected with the strained BTO with a continuous tail, while the one grown on paraelectric PTO is separated from the strained part. The reciprocal space maps can be assembled into continuous movies that allow the observation of the relaxation process during the growth (Supplementary Movie 1, 2). 

A role for polarization in influencing strain relaxation can also be inferred from Landau theory. We again impose the initial boundary condition for BTO growing on PTO that the polarization is continuous across the interface, take the calculated polarization for the thin PTO layer $P_{PTO}$ as the polarization of BTO and calculate $\frac{\partial G}{\partial u_{m}}$ for BaTiO$_{3}$.

\begin{equation}
\frac{\partial G}{\partial u_{m}}=-\frac{2Q_{12}}{s_{11}+s_{12}}P_{PTO}^{2}+\frac{2u_{m}}{s_{11}+s_{12}}
\end{equation}

In the above equation the elastic constants are those of BTO, whereas $P_{PTO}$ is calculated from Equation 1 using the appropriate coefficients for PTO. Under all of the experimental conditions $\frac{\partial G}{\partial u_{m}}$ is negative, implying a driving force towards positive strain, ie. relaxation of the negative misfit strain induced by the substrate. However it is indeed seen that the polarization of the PTO layer reduces this driving force by making the term  $\frac{\partial G}{\partial u_{m}}$ more positive. $\frac{\partial G}{\partial u_{m}}$ is plotted as function of PTO thickness and growth temperature in Fig. \ref{RSM_img} (c).  

\begin{figure}[ht!]
	\begin{center}
	\includegraphics[width=10cm]{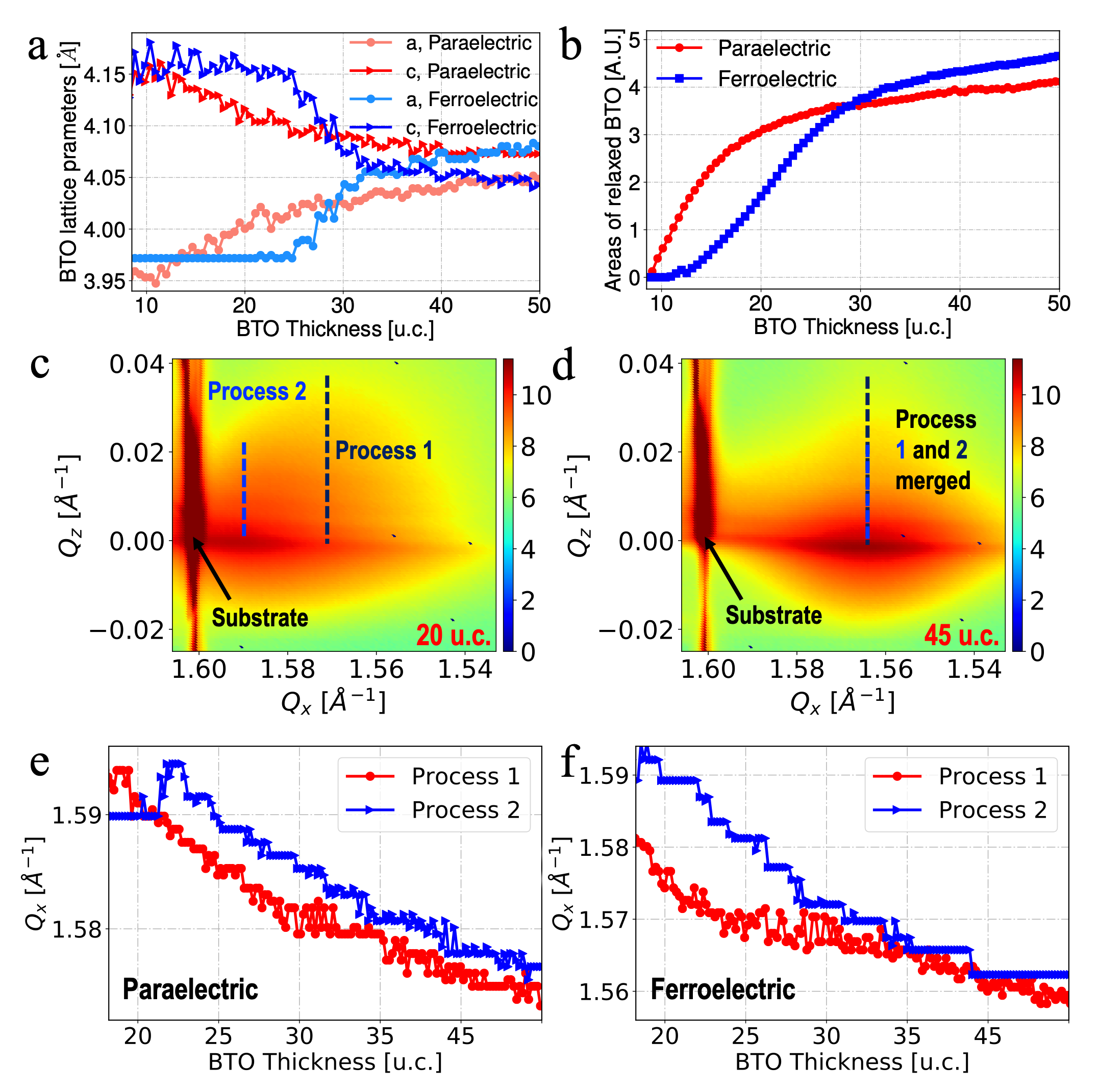}
		
		\caption[Relaxation process]{(a-b) Relaxation process: 
			(a) In-plane and out-of-plane lattice parameters of relaxed BTO plotted against BTO growth time for two samples: one grown on paraelectric PTO (3 u.c. PTO, 565\textdegree C, red and light red represent lattice parameters c and a, respectively), and another grown on ferroelectric PTO with largest polarization among all samples (4 u.c. PTO, 500$^{\circ}$ C, blue and light blue represent lattice parameters c and a, respectively). Source data are provided as a Source Data file. (b) Areas of the relaxed part of BTO film plotted against BTO growth time for two samples described in (a): red grown on paraelectric PTO and blue grown on ferroelectric PTO. Source data are provided as a Source Data file. (c,d) Examples of reciprocal space maps obtained by in-situ grazing incidence diffraction in the vicinity of the (1\,0\,0) substrate peak (seen on the left hand side of the map at $Q_{x}\sim 1.60$) for BTO films grown on 6 layers of PTO films at 565$^{\circ}$ C after (a) 20 u.c. layers and  (b) 45 u.c. layers BTO were grown. Two types of relaxation process were seen which begin at different $Q_x$ positions in the beginning of growth and merge to the same $Q_x$ positions at the end of the growth. (e, f) Evolution of the $Q_x$ positions of the two relaxation features plotted against BTO thickness for two samples: one grown on paraelectric PTO ((c), 3 u.c. PTO, 580$^{\circ}$ C), and another grown on ferroelectric PTO with largest polarization among all samples ((d), 3 u.c. PTO, 500$^{\circ}$ C). Red and blue data represent Process 1 and Process 2 respectively. Source data are provided as a Source Data file.}
		\label{relaxation_graphs}
	\end{center}
\end{figure}

The lattice parameters of relaxed BTO obtained from the center of relaxed parts are plotted against growth time for two representative samples in Fig. \ref{relaxation_graphs} (a). All the films initially grow constrained in-plane to  the substrate. Once relaxation begins the a and c lattice parameters start to  evolve towards each other and eventually match bulk BTO lattice parameters. The BTO films grown on PTO with larger polarization stay in the strained ferroelectric state longer and starts relaxation later, but arrives to bulk lattice parameters earlier, implying a quicker relaxation process once it begins. The areas of the relaxed BTO were calculated by adding up all pixels above a certain limit in the relaxed region and are plotted against BTO thickness in (Fig. \ref{relaxation_graphs} (b)). Two different relaxation process of films grown on ferroelectric/paraelectric ``substrates'' can be seen.  

Further insight into the difference in the in-plane relaxation process can be obtained by performing grazing-incidence X-ray scattering around the (1\,0\,0) peak. Examples from the beginning and end of growth for sample grown on a ferroelectric substrate are shown in Fig. \ref{relaxation_graphs} (c, d). A full set of images for the growth of a sample on a ferroelectric substrate and a paraelectric substrate are shown in Supplementary Fig. 9 and 10 respectively. Two relaxation processes are evident in this data. Process 1 has a larger  spread in $Q_z$ and while Process 2 has a much smaller spread in $Q_z$. On ferroelectric PTO (Fig. \ref{relaxation_graphs} (f)) these two process occur at different $Q_x$ positions at the beginning of growth and merge to the same $Q_x$ positions at the end of the growth (this also can be seen by comparing (c) and (d)). By contrast the two processes are parallel from the beginning to the end of the growth if the film is grown on paraelectric PTO (Fig. \ref{relaxation_graphs} (e)). Process 2 appears to occur through jumps between intermediate states where $Q_x$ remains constant for a while during relaxation (blue curve in (e) and (f)). These jumps are larger and happen less frequently in the case of samples grown on ferroelectric substrates, which is compatible  with our argument that the polarization essentially reduces the free energy benefit of strain relaxation and thereby increases the activation energy required for relaxation events to occur.

In conclusion, by combining results from XRD, PFM, electrical measurements, and especially in-situ XRD at NSLS-II, we have demonstrated that the ferroelectric polarization underneath BTO films during growth can help them stay in highly-strained states. As an example, if we consider a device such as a ferroelectric memory based on a capacitor or ferroelectric field effect transistor, films below 10nm in thickness are not particularly useful due to increased leakage currents\cite{DRS05} and suppressed polarization due to depolarization fields\cite{Junquera03}. On the other hand the relaxation of strain as thickness increases leads to lower useful polarization in relaxed films. At a thickness of 10nm ($\sim 25$ unit cell layers) a film grown on a paraelectric substrate has lattice parameters near to bulk, whereas one grown on a ferroelectric substrate is completely strained (Fig. \ref{relaxation_graphs} (a)), showing that our method could have real implications at a thickness of practical importance. In other applications, such as domain wall electronics, where the conductive properties of domain walls can be used as the basis of nanoelectronics\cite{Seidel09,Guyonnet11,Maksymovych11,Farokhipoor11,Catalan12,Sluka13,Whyte15} the density and arrangement of as-grown domains is important, and this also can be modified through our method. Since the polarization during growth is a parameter that can be easily and widely adjusted via growth conditions such as temperature and thickness of the underlying ferroelectric layer there is additional flexibility compared to strain engineering, which is limited by the availability of appropriate substrates.  Ferroelectric properties, domain configuration, and the strain state of the thin film can all be manipulated via the underlying ferroelectric polarization of the PTO film during growth, making polarization engineering a powerful approach to the design of tailored ferroelectric films for specialized applications.

\newpage

\section*{Methods}

\subsection{Synthesis of films using off-axis RF magnetron sputtering.}
Bilayers films of BTO/PTO  were grown on 20nm SrRuO$_{3}$/SrTiO$_3$ substrates via off-axis RF magnetron sputtering.  The SrRuO$_{3}$ electrodes were grown at a pressure of 0.1 Torr, Ar:O of 16:3 and a growth temperature of 610$^{\circ}$ C. The growth conditions used for the deposition of PTO and BTO were exactly the same: a pressure of 0.18 Torr, Ar:O ratio of 16:7 and  the same temperature was used for both films. The growth temperature of the samples spanned the range from 475$^{\circ}$ C to 580$^{\circ}$ C. The bottom 20nm thick SrRuO$_{3}$ electrodes were grown prior to the in-situ experiments in the off-axis magnetron sputtering chamber at Stony Brook University. The SrRuO$_{3}$ electrodes were also atomically flat with single unit cell steps which were checked by Atomic force microscopy. X-ray diffraction results prior to deposition show that these films were epitaxially constrained to the SrTiO$_3$ substrates and had the same in-plane lattice parameter as the substrates.

\subsection{Growth rate calibration using fitting method}
The growth time per BTO layer was determined by fitting the $\theta-2\theta$ scans and the low angle reflectivity scans using the same fitting method as in \cite{Bein15}. We can determine the total thickness of film by fitting the low angle reflectivity scans, and the thickness of each material (with different tetragonality) by fitting the  $\theta-2\theta$ scans around (0\,0\,1) and (0\,0\,2) Bragg peak. All these fitted results confirmed each other to obtain more accurate growth rates. The growth rates at 4-ID beamline was also determined by in-situ measurement at the anti-Bragg peak (Supplementary Fig. 6).

\subsection{Atomic force and piezoelectric force microscopy}
The films were characterized using an atomic force microscope (MFP-3D, Asylum Research) to access the morphology. Vector PFM was conducted to study the ferroelectric domain structure both in-plane and out-of-plane simultaneously. For PFM measurements, conductive Co/Cr-coated silicon tips (spring constant 2.8 N/m) was used and an alternating 0-5 V voltage was applied between the tip and the SrRuO$_{3}$ bottom electrode.

\subsection{Details about in-situ x-ray diffraction set-up}
The in-situ x-ray diffraction experiments presented here were performed at the NSLS-II 4-ID beamline. At beamline 4-ID, a 2.8m long undulator is the source of photons, and a monochromator  with Si(111) crystals selects $\sim 2\times10^{12}$ photons per second at a photon energy of 11.42 keV with an energy bandwidth of $\sim 2\times10^{4}$ from that source. Mirrors located approximately midway down the beamline with a Pd coating were used to refocus the photons onto the surface of the sample. Beam stability is enhanced with a feedback loop consisting of a diamond beam position detector and piezo actuator for angular adjustment of the second monochromator crystal. The experiments were performed in an in-situ growth chamber with temperature, pressure and atmospheric control. Four angles $\phi, \theta, \delta$ and $2\theta$ can be controlled by computer during the experiments. The x-ray beam enters the chamber through a beryllium window, scatters off the sample, exits via a second beryllium window and is detected by an Eiger 1M area detector. In the grazing incidence experiment, the sample is inclined by a small grazing angle with respect to the incoming X-ray beam. 

\subsection{In-situ x-ray diffraction data analysis}

The in-situ x-ray diffraction data obtained at 4-ID beamline were saved in HDF5 files. A methodology was developed using Python to analyze the files. The 2D array data were converted to 3D data with angles and reciprocal space parameters to perform reciprocal space maps. The reciprocal space growth movies were generated by merging all the maps in the order of time. The x-ray intensity of different samples were normalized to be comparable. The relaxed area of BTO films were defined by setting up a certain threshold for all samples after normalization. The positions of multipeaks were obtained using several methods to reduce errors.

\section*{Acknowledgments}
This work was supported by NSF DMR-1055413, DMR-1334867 and DMR-1506930. Development of the in situ growth facility used in this work was supported by NSF DMR-0959486. This research used beamline 4-ID of the National Synchrotron Light Source II, a U.S. Department of Energy (DOE) Office of Science User Facility operated for the DOE Office of Science by Brookhaven National Laboratory under Contract No. DE-SC0012704. We thank C. Nelson and Z. Yin for assistance with the experiments at 4-ID.

\section*{Author Contributions}
R.L., J.G.U., H.-C.H, A.G., B.B, G.B., A.L., K.E.-L., R.L.H and M.D  participated in the experimental work at NSLS-II.  R.L. conducted the bulk of the experimental work outside of NSLS-II and supervised A.S., C.P., K.C. and E.D who completed high school and undergraduate research projects related to the research. R.L. was chiefly responsible for processing and analysis of the data. The experimental apparatus was designed, assembled and integrated in to the 4-ID beamline by R.L.H. and M.D.. M.D. devised the experimental methods and directed the project. The manuscript was prepared by R.L. and M.D. with contributions from the other authors.

\newpage

\section*{Supplementary Information for ``Role of ferroelectric polarization during growth of highly strained ferroelectrics revealed by in-situ x-ray diffraction'' by Rui Liu et al.}
\setcounter{page}{1}

\setcounter{subsection}{0}
\setcounter{figure}{0}

\subsection{Structural characterization using laboratory-based X-ray diffraction.}
Crystal structures of the films were
determined by X-Ray Diffraction on a Bruker D8-Discover high-resolution X-Ray Diffractometer. The wavelength used was 1.5406 \AA. Here we show three BaTiO$_{3}$ films grown on top of 3 unit cell PbTiO$_{3}$ ultra-thin films as examples. The growth temperatures are below, around or above the ferroelectric transition temperature of 3 unit cell PbTiO$_{3}$ ultrathin film (550$^{\circ}$ C). All samples showed epitaxial growth and the out of plane lattice parameters are close to each other (See Supplementary Fig. \ref{XRD}).

\begin{figure*}[htbp]
	\begin{center}
		\includegraphics[width=16cm]{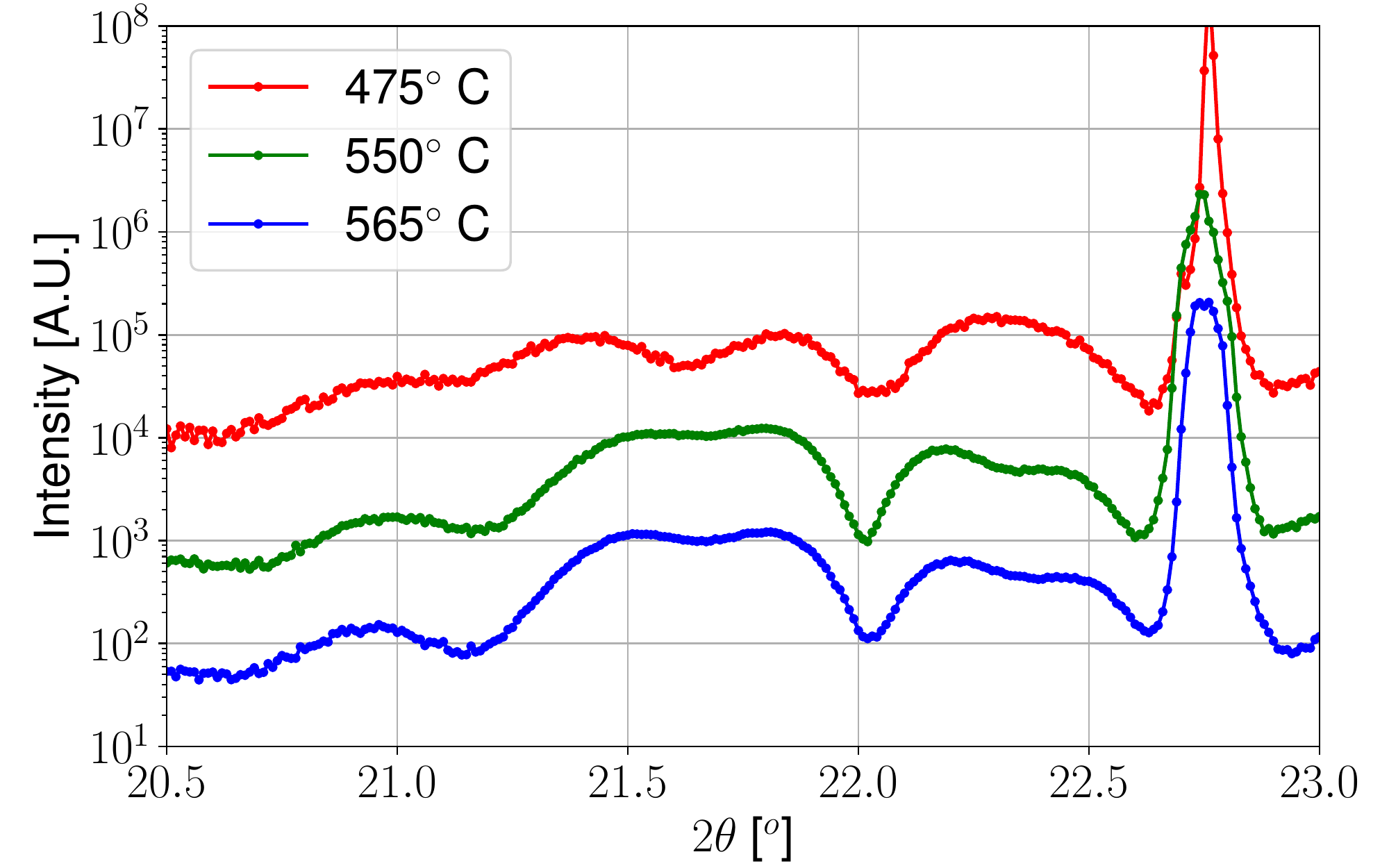}
		
		\caption[Examples of $\theta-2\theta$ scans around (0\,0\,1) Bragg peak]{Examples of $\theta-2\theta$ scans around (0\,0\,1) Bragg peak: 
			BaTiO$_{3}$ films grown on 3 unit cell PbTiO$_{3}$ films at growth temperatures of 475$^{\circ}$ C (red), 550$^{\circ}$ C (green), 565$^{\circ}$ C (blue). The growth temperatures are below (red), around (green) or above (blue) the ferroelectric transition temperature of 3 unit cell PbTiO$_{3}$ film. Source data are provided as a Source Data file.}
		\label{XRD}
	\end{center}
\end{figure*}

\subsection{Topography characterization.}
The topography of films was characterized using an atomic force microscope (MFP-3D, Asylum Research). The surfaces of all the films were clean and atomically flat with single unit cell steps of 0.4 nm (See Supplementary Fig. \ref{AFM}).
\begin{figure*}[htbp]
	\begin{center}
		\includegraphics[width=16cm]{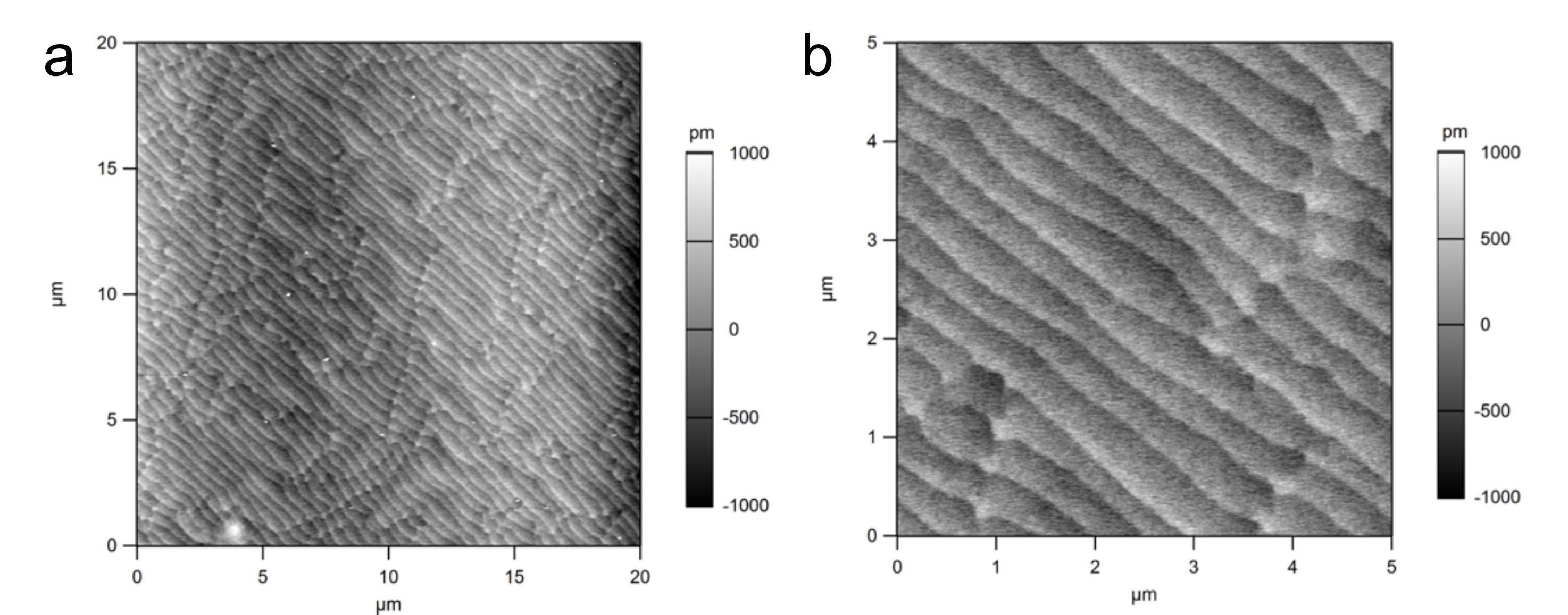}
		
		\caption[Examples of topography of BaTiO$_{3}$ films]{Examples of topography of BaTiO$_{3}$ films:
			AFM topography of BaTiO$_{3}$ film grown on 3 unit cell of PbTiO$_{3}$ films with scan sizes (20 $\times$ 20) $\mu m^{2}$ and (5 $\times$ 5) $\mu m^{2}$.}
		\label{AFM}
	\end{center}
\end{figure*}

\subsection{Ferroelectric polarization measurements using PUND.}
Ferroelectric polarization was measured using a PUND technique. PUND, or Positive Up Negative Down, is a pulse train that uses repeating up and then repeating down pulses to first switch the polarization and then gather information about non-switching currents from the second pulse. Using this technique any non-switching contributions to the current were subtracted out. Polarizations of three BaTiO$_{3}$ films grown on 3 unit cell of PbTiO$_{3}$ films at different temperatures were presented in Supplementary Fig. \ref{PUND}, inserted with switching currents collected during the PUND measurements. Multiple measurements were conducted for each sample to obtain accurate results with statistical significance.
\begin{figure*}[htbp]
	\begin{center}
		\includegraphics[width=16cm]{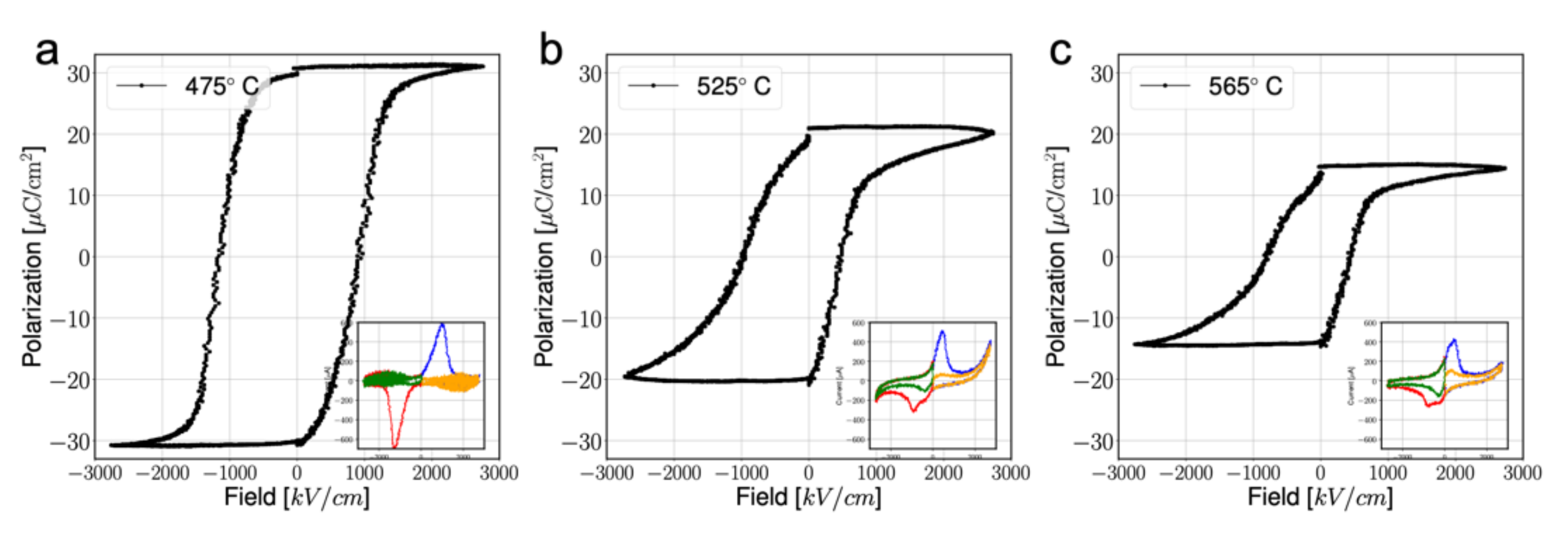}
		
		\caption[Examples of hysteresis loops]{Examples of ferroelectric hysteresis loops measured using a PUND technique: 
			Ferroelectric hysteresis loops of BaTiO$_{3}$ films grown on 3 unit cell of PbTiO$_{3}$ films at growth temperatures of 475$^{\circ}$ C (a), 525$^{\circ}$ C (b), 565$^{\circ}$ C (c). The inserted figure at the right bottom corner of each figure was the switching currents collected in the PUND measurement versus the applied electrical field. Color guide: P pulse (blue), U pulse (orange), N pulse (red), D pulse (green).  Source data are provided as a Source Data file.}
		\label{PUND}
	\end{center}
\end{figure*}

\subsection{Electrical properties of the films.}

To measure the piezoelectric response of the samples DART (Dual Amplitude Resonance Tracking) Piezo-Force Microscopy (PFM) measurements were made using Co/Cr-coated AFM tip as the top electrode. The measured amplitudes and phases response recorded during the measurement is shown in Supplementary Fig. \ref{Electrical Measurements} (a,b). It can be seen that all the samples are very good ferroelectrics at room temperature and they have similar piezoelectric coefficients d$_{33}$. The dielectric constant was also measured as a function of applied voltage using an LCR circuit and all samples displayed the expected butterfly loop characteristic of strongly ferroelectric materials (See Supplementary Fig. \ref{Electrical Measurements} (c)). All the films have similar dielectric constants between 150 to 200.

One very intriguing observation we have made is that while the magnitude of piezoelectric coefficients is similar for all of the films, those films were grown on ferroelectric PbTiO$_{3}$ substrates have an electromechanical resonance frequency about 10 \% lower compared to those grown on paraelectric substrates (See Supplementary Fig. \ref{PFM freq}). The ferroelectric polarization and the surface domain configuration can affect the elastic properties of the samples, which in turn had an influence on the electromechanical resonance frequency.

\begin{figure*}[htbp]
	\begin{center}
		\includegraphics[width=16cm]{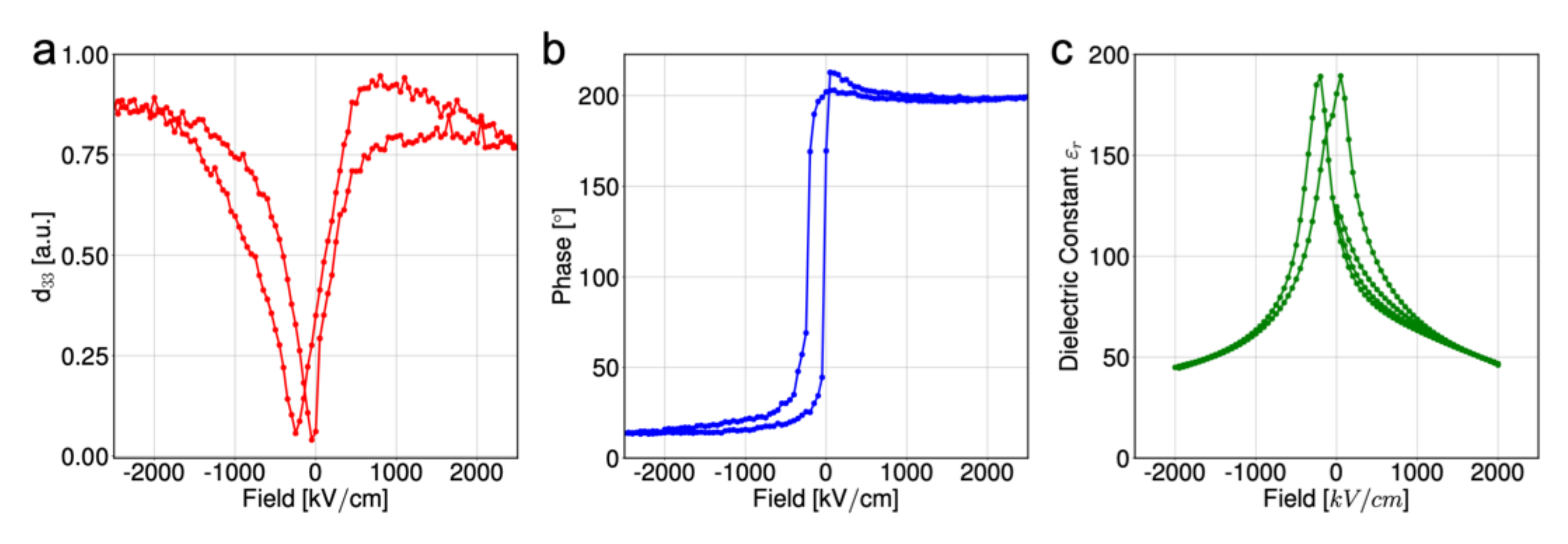}
		
		\caption[Electrical properties of the films]{Electrical properties of the films:
			(a,b) The measured amplitudes and phases (a and b, respectively) of the tip response during the DART PFM measurements were plotted against the applied field. (c) The dielectric constant was also measured as a function of applied field. The butterfly loop characteristic showed strong ferroelectricity.  Source data are provided as a Source Data file.}
		\label{Electrical Measurements}
	\end{center}
\end{figure*}

\begin{figure*}[htbp]
	\begin{center}
		\includegraphics[width=16cm]{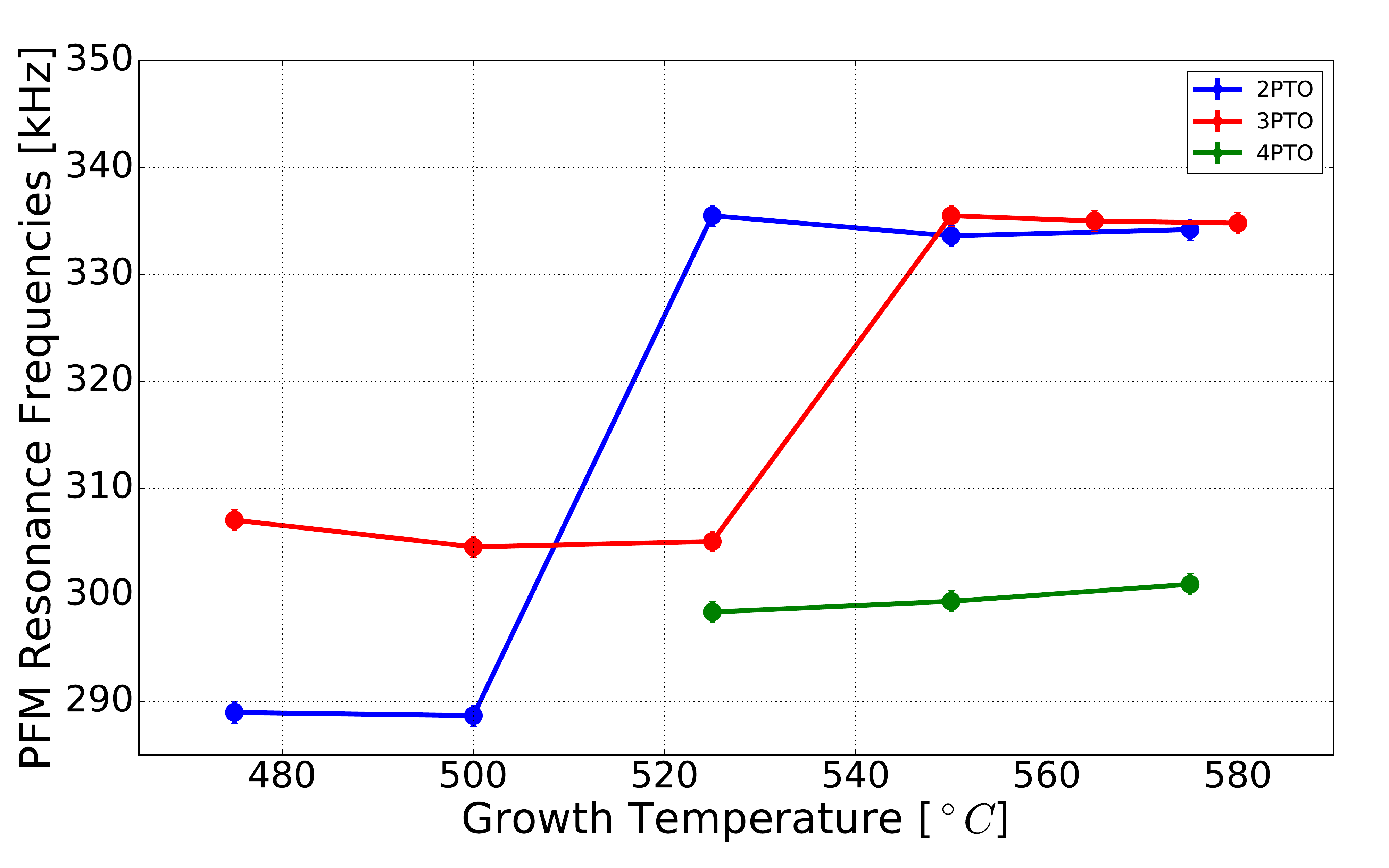}
		
		\caption[Electromechanical resonance frequency]{Electromechanical resonance frequences of samples measured by DART PFM: 
			Electromechanical resonance frequences of BaTiO$_{3}$ films grown on 2 (blue), 3 (red), 4 (green) unit cell PbTiO$_{3}$ films versus growth temperatures.  Source data are provided as a Source Data file.}
		\label{PFM freq}
	\end{center}
\end{figure*}

\subsection{Growth monitor using the anti-Bragg peak.}
To monitor the growth and also to calibrate the growth rates, the intensity of the reflected signal at the (0\,0\,$\frac{1}{2}$) Bragg position (to minimize bulk Bragg diffraction) was measured as a function of growth time. The oscillations of the reflectivity signal provide a measurement of surface roughness similar to reflection high energy electron diffraction (RHEED). While the RHEED technique is not appropriate here due to the presence of magnetic fields and does not provide the same amount of structural information as x-ray diffraction can. The maxima intensity correspond to the completed layers while the low signals correspond to incomplete layers. The oscillations indicate that the BTO grown on paraelectric substrate did not maintain a smooth surface, while BTO grown on ferroelectric PTO continues to grow smoothly in a layer-by-layer mode for many layers (See Supplementary Fig. \ref{Ocillation}).  The ferroelectric polarization of PTO underneath helps maintains the BTO growth in a smooth layer-by-layer mode.

\begin{figure*}[htbp]
	\begin{center}
		\includegraphics[width=16cm]{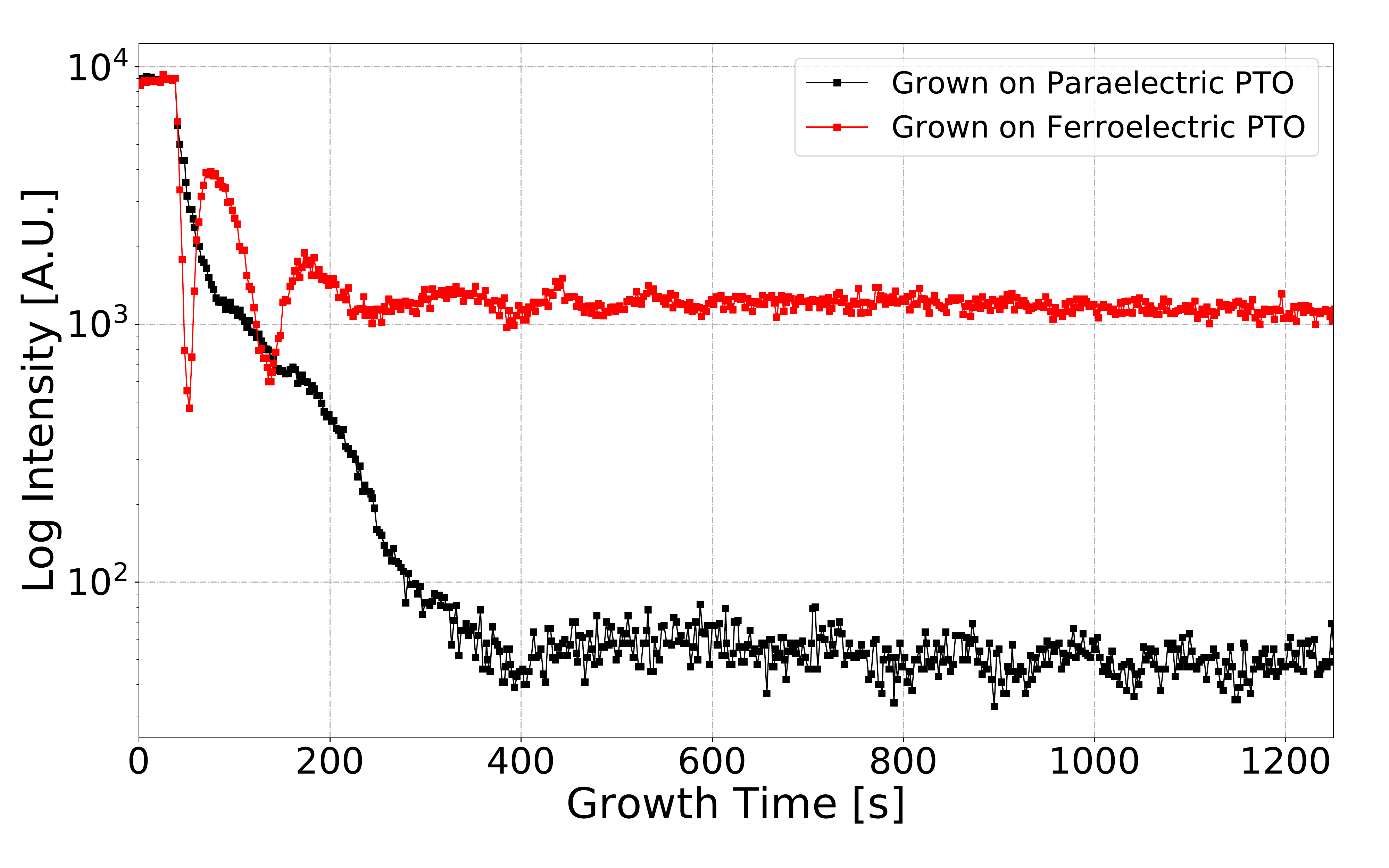}
		
		\caption[Growth monitor using the anti-Bragg peak.]{Growth monitor using the anti-Bragg peak: 
			Intensity of the (0\,0\,$\frac{1}{2}$) anti-Bragg peak versus growth time for BaTiO$_{3}$ films grown on paraelectric (black) and ferroelectric (red) PbTiO$_{3}$ films. It can be seen that the
			BTO grown on paraelectric PTO did not grow in a smooth way, while BTO grown on ferroelectric PTO grows smoothly in the first few layers.}
		\label{Ocillation}
	\end{center}
\end{figure*}

\subsection{In-situ X-ray scans near (1\, 0\, 1\,) peak.}

The evolution of the relaxation of BTO can be obtained from the scans near  the (1\, 0\, 1\,) peak (Two extreme examples were presented in the paper). All the films grow strained with the substrate at the beginning of growth, then the lattice parameters a and c start to merge at different thickness and tend toward bulk lattice parameters in the end. (See Supplementary Fig. \ref{ISR 101} (a,b)). The area of the relaxed BTO was calculated by adding up all pixels above a certain limit in the relaxed region (See Supplementary Fig. \ref{ISR 101} (c,d)).

\begin{figure*}[htbp]
	\begin{center}
		\includegraphics[width=16cm]{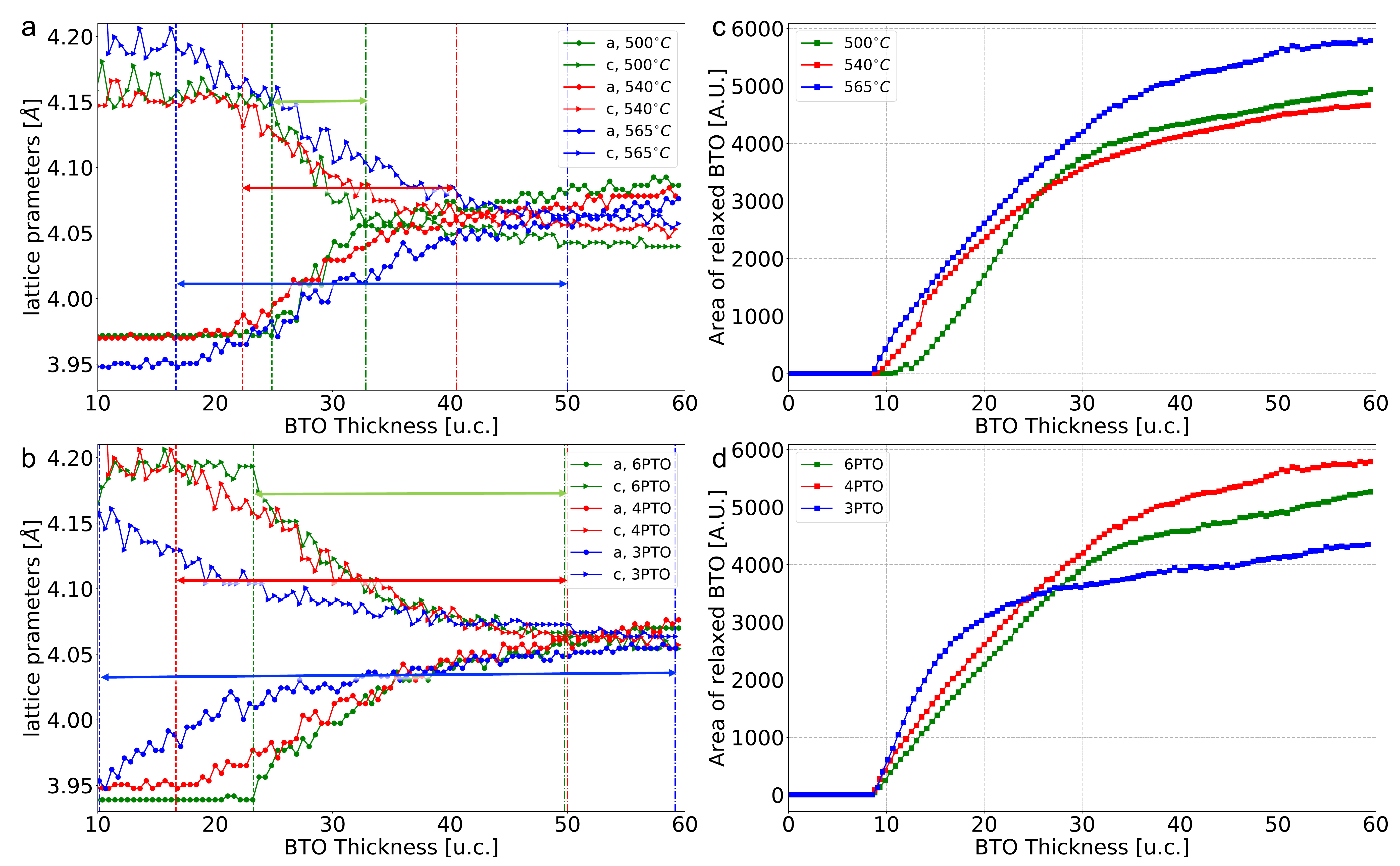}

		\caption[Relaxiation process presents by lattice parameters and size of the relaxed part]{Evolution of lattice parameters and sizes of relaxed parts of BTO films obtained by in-situ x-ray scans near (1\, 0\, 1\,) peak: 
				(a,b) In-plane lattice parameters a (circle date point markers) and out of plane lattice parameters c (triangle data point markers) of relaxed BTO were plotted against BTO thickness for two series of samples: one keeps the unit cell layers of PTO films to be 4 while varying the growth temperatures (a, green: 500$^{\circ}$ C, red: 540$^{\circ}$ C, blue: 565$^{\circ}$ C), and another keeps the growth temperatures to be 565$^{\circ}$ C while varying unit cell layers of PTO films (b, green: 6 layers, red: 4 layers, blue: 3 layers). The vertical dashed line indicates the beginning of relaxation and the vertical dash-dotted line indicates the end of relaxation for each film. The horizontal arrow points out the whole range of the relaxation process. (c,d) Area of relaxed part of BTO films plotting against BTO thickness for two series of samples: one keeps the unit cell layers of PTO films to be 4 while varying the growth temperatures (a, green: 500$^{\circ}$ C, red: 540$^{\circ}$ C, blue: 565$^{\circ}$ C), and another keeps the growth temperature to be 565$^{\circ}$ C while varying unit cell layers of PTO films (b, green: 6 layers, red: 4 layers, blue: 3 layers). Source data are provided as a Source Data file.}
		\label{ISR 101}
	\end{center}
\end{figure*}

\subsection{Reciprocal space growth movies obtained by in-situ x-ray scans around (1\, 0\, 1\,) peak.}
The reciprocal space maps can be assembled into continuous movies that allow the observation of the growth process (Supplementary Movie 1). Two series of samples were grown: one keeps the thickness of PTO films to be the same while varying the growth temperatures, and another keeps the growth temperatures to be the same while varying the thickness of PTO films. The polarization of PTO substrates was manipulated by both ways (see the inserted image on top). Epitaxially strained growth was observed during the whole process, while the BTO films start partially relaxing at different growth time. In general, the ferroelectric polarization underneath can help constrain BTO films to the lattice parameters of the substrates, maintaining large compressive strain in the film. The relaxation process was delayed by the polarization of the substrates and the relaxation mode was also changed. 

\subsection{Analysis of domains from in-situ x-ray data around (0\, 0\, 1\,) peak.}
The evolution of domains during the growth can be obtained from in-situ x-ray scans around (0\, 0\, 1\,) peak. First, we integrated the intensity along the $Q_z$ direction around the Bragg peak of BTO. Then we fitted the Bragg peak plus two domain peaks using three independent Lorentz functions along the $Q_x$ direction (See Supplementary Fig. \ref{Fitting}). Finally, we calculated the domain size from the distances between the Bragg peak and the Domain peaks. Several fitting methods were used to cross-check the results. After applying the method to every scan of all films, we obtained the evolution of domain sizes for all samples.

\begin{figure*}[htbp]
	\begin{center}
		\includegraphics[width=16cm]{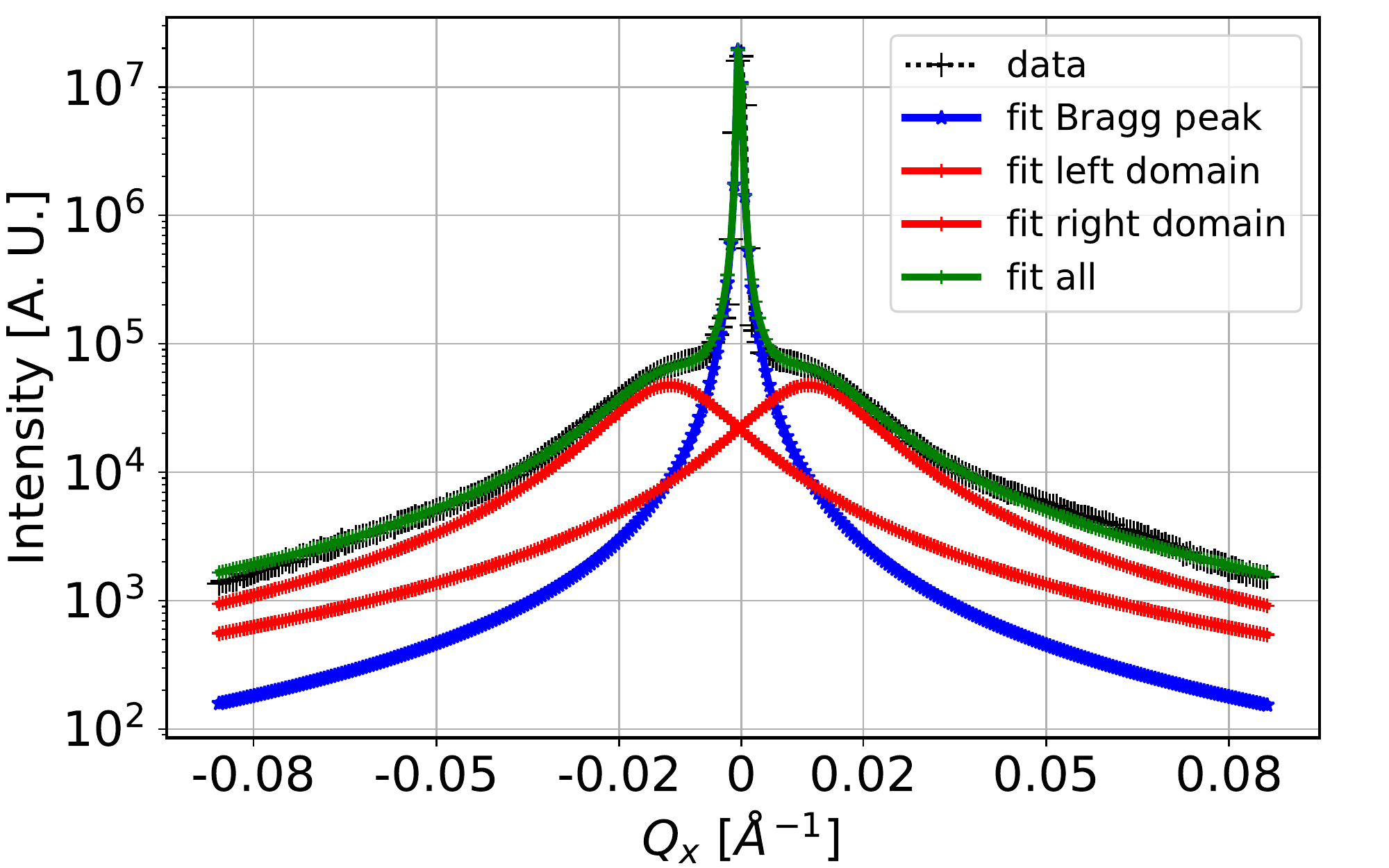}
		
		\caption[An example of domain fitting.]{An example of domain fitting: 
			The integrated plot around the Bragg peak of BTO along the Qx line together with a fit (green) to the data (black). The Bragg peak (blue) and the domain diffuse  (red) were fitted using three independent Lorentz functions.}
		\label{Fitting}
	\end{center}
\end{figure*}

\subsection{Reciprocal space growth movies obtained by in-situ x-ray scans around (0\, 0\, 1\,) peak.}
The reciprocal space growth movies around (0\, 0\, 1\,) peak (Supplementary Movie 2) were obtained using the same method as (1\, 0\, 1\,) growth movies. The polarization of PTO substrates was also inserted in the movie.

\subsection{Grazing-incidence in-situ x-ray scans near (1\, 0\, 0\,) peak.}

Further insight into the difference in the in-plane relaxation process can be obtained
by performing grazing-incidence X-ray scattering around the (1 0 0)
peak. Examples from the beginning to the end of growth of BTO film grown on ferroelectric and paraelectric PTO substrates are shown in Supplementary Fig. \ref{ISR 100 ferro} and Supplementary Fig. \ref{ISR 100 para}. Two types of relaxation process were observed: Process 1 (around the green dashed line in Supplementary Fig. \ref{ISR 100 ferro} (c,f)) with larger $Q_z$ and Process 2 (around the blue dashed line in Supplementary Fig. \ref{ISR 100 ferro} (c,f)) with smaller $Q_z$. These two processes happened at different $Q_x$ positions at the beginning of growth and merged to the same $Q_x$ positions at the end of the growth if grown on the ferroelectric substrate (Supplementary Fig. \ref{ISR 100 ferro}). By contrast in the samples grown on paraelectric substrates the two processes are at the same $Q_x$ positions during the whole growth process (Supplementary Fig. \ref{ISR 100 para}).

\begin{figure*}[htbp]
	\begin{center}
		\includegraphics[width=16cm]{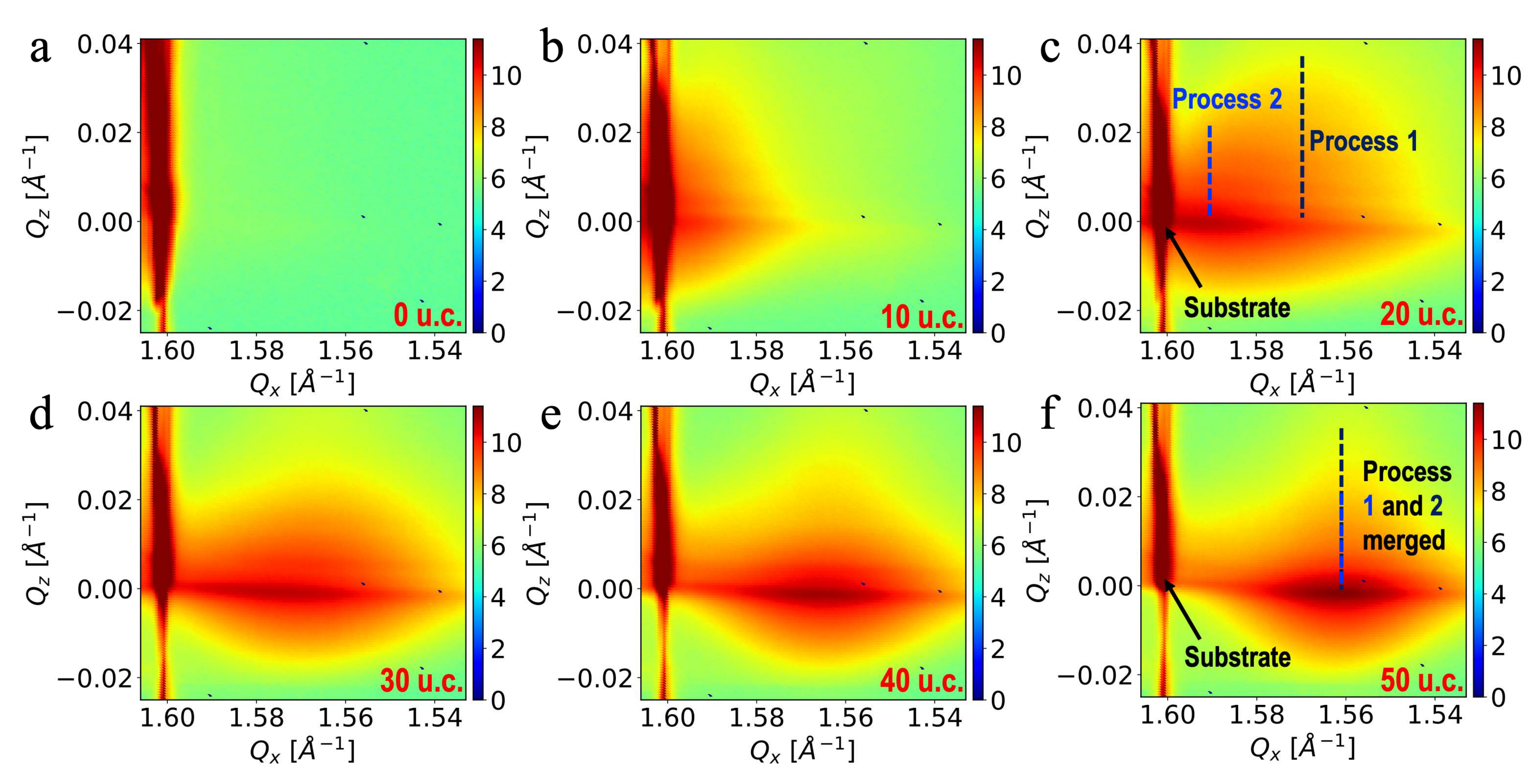}		
		
		\caption[Grazing-incidence in-situ x-ray scans near (1\, 0\, 0\,) peak of film grown on ferroelectric substrate]{Examples of reciprocal space maps obtained by in-situ Grazing-incidence x-ray scans around (1 0 0) peak for BTO films grown on ferroelectric substrate after (a) 0 u.c., (b) 10 u.c., (c) 20 u.c., (d) 30 u.c., (e) 40 u.c., and (f) 50 u.c. BTO films were grown. Two types of relaxation processes were seen which begin at different $Q_x$ positions at the beginning of growth and merge to the same Qx positions at the end of the growth.}
		\label{ISR 100 ferro}
	\end{center}
\end{figure*}

\begin{figure*}[htbp]
	\begin{center}
		\includegraphics[width=16cm]{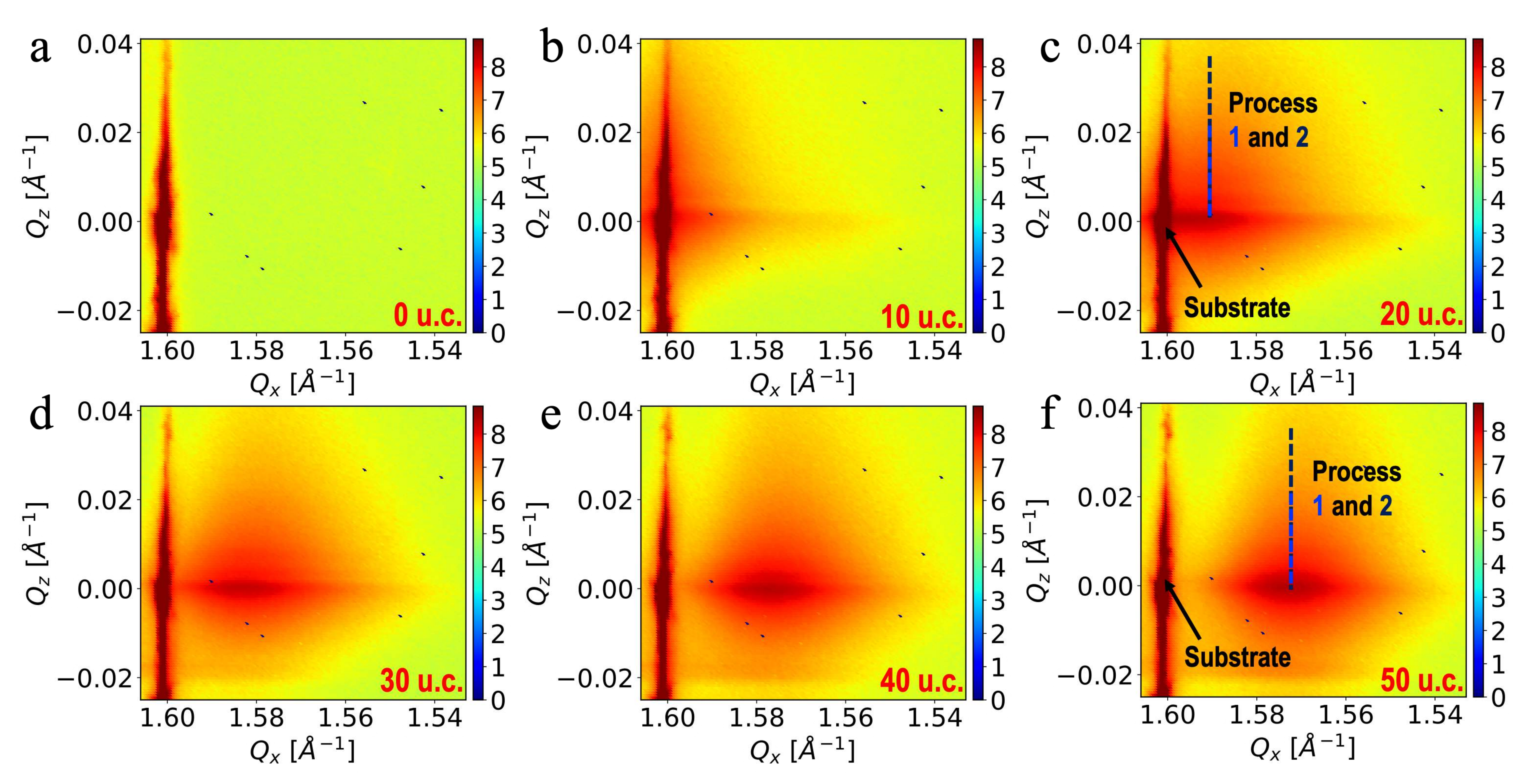}

		\caption[Grazing-incidence in-situ x-ray scans near (1\, 0\, 0\,) peak of film grown on paraelectric substrate]{Examples of reciprocal space maps obtained by in-situ Grazing-incidence x-ray scans around (1 0 0) peak for BTO films grown on paraelectric substrate after (a) 0 u.c., (b) 10 u.c., (c) 20 u.c., (d) 30 u.c., (e) 40 u.c., and (f) 50 u.c. BTO films were grown. The two processes are at the same $Q_x$ positions during the whole growth process.}
        \label{ISR 100 para}
	\end{center}
\end{figure*}


\newpage


\begin{thebibliography}{99}
	
	
	
\bibitem{Choi04}
Choi, K. J. et al. Enhancement of ferroelectricity in strained BaTiO3 thin films. \emph{Science} \textbf{306} 1005-1009 (2004)
	
\bibitem{Bea09}
Bea, H. et al. Evidence for Room-Temperature Multiferroicity in a Compound with a Giant Axial Ratio. \emph{Phys. Rev. Lett.} \textbf{102} 217603 (2009).
	
\bibitem{Lee10}
Lee, J. H. et al. A strong ferroelectric ferromagnet created by means of spin-lattice coupling. \emph{Nature} \textbf{466} 954-U972 (2010)
	
\bibitem{Ryan13}
Ryan, P. J. et al. Reversible control of magnetic interactions by electric field in a single-phase material. \emph{Nat. Commun.} \textbf{4} 1334 (2013).
	
\bibitem{Junquera03}
Junquera, J. \& Ghosez, P. Critical thickness for ferroelectricity in perovskite ultrathin films. \emph{Nature} \textbf{422} 506-509 (2003).
	
\bibitem{Dawber03}
Dawber, M., Chandra, P.,  Littlewood, P.B. \& Scott, J.F. Depolarization corrections to the coercive field in thin-film ferroelectrics. \emph{J. Phys. Condens. Mater.} \textbf{15} L393-L398 (2003).
	
\bibitem{Lichtensteiger05}
Lichtensteiger, C., Triscone, J.-M., Junquera, J. \& Ghosez, P. Ferroelectricity and tetragonality in ultrathin PbTiO$_{3}$ films. \emph{Phys. Rev. Lett} \textbf{94} 047603 (2005).
	
\bibitem{Streiffer02}
Streiffer, S.K. et al. Observation of nanoscale 180 degrees stripe domains in ferroelectric PbTiO3 thin films. \emph{Phys. Rev. Lett} \textbf{89} 067601 (2002).
	
\bibitem{Fong04}
Fong, D.D. et al. Ferroelectricity in ultrathin perovskite films. \emph{Science} \textbf{304} 1650-1653 (2004).
	
\bibitem{Fong06}
Fong, D.D., et al., Stabilization of Monodomain Polarization in Ultrathin PbTiO3 Films. \emph{Phys. Rev. Lett.} \textbf{96} 127601 (2006).
	
	\bibitem{Dawber05}
Dawber, M. et al. Unusual behavior of the ferroelectric polarization in PbTiO$_3$/SrTiO$_3$ superlattices. \emph{Phys. Rev. Lett} \textbf{95} 177601 (2005).

	\bibitem{Dawber07}
Dawber, M. et al. Tailoring the properties of artificially layered ferroelectric superlattices. \emph{Adv. Mater.} \textbf{19} 4153 (2007).

\bibitem{McQuaid11}
McQuaid, R.G.P., McGilly, L.J., Sharma, P., Gruverman, A. \& Gregg J.M. Mesoscale flux-closure domain formation in single-crystal BaTiO3. \emph{Nature Comm.} \textbf{2} 404 (2011).
	
\bibitem{Zubko12}
Zubko, P., Jecklin, n., Torres-Pardo, A., Aguado-Puente, P., Gloter, A., Lichtensteiger, C., Junquera, J., Stephan, O. \& Triscone, J.-M. Electrostatic Coupling and Local Structural Distortions at Interfaces in Ferroelectric/Paraelectric Superlattices. \emph{Nanotletters} \textbf{12} 2846 (2012). 
	
\bibitem{Lichtensteiger14}
Lichtensteiger, C., Fernandez-Pena, S., Weymann, C., Zubko, P. \& Triscone, J.-M. Tuning of the Depolarization Field and Nanodomain Structure in Ferroelectric Thin Films. \emph{Nanoletters} \textbf{14} 4205 (2014).
	
\bibitem{Bein15} 
Bein, B. et al. In-situ x-ray diffraction and the evolution of polarization during the growth of ferroelectric superlattices \emph{Nat. Commun.} \textbf{6} 10136 (2015).
	
\bibitem{Lichtensteiger16}
Lichtensteiger, C., Weymann, C., Fernandez-Pena, S., Paruch, P. \& Triscone, J.-M. Built-in voltage in thin ferroelectric PbTiO$_{3}$ films: the effect of electrostatic boundary conditions. \emph{New Journal of Physics} \textbf{18} 043030 (2016).

	
\bibitem{Yadav16}
Yadav, A.K. et al. Observation of polar vortices in oxide superlattices \emph{Nature} \textbf{530} 198 (2016)

\bibitem{Hong17}
Hong, Z. et al. Stability of Polar Vortex Lattice in Ferroelectric Superlattices \emph{Nano Lett.} \textbf{17} 2246 (2017).

\bibitem{Damoradan17}
Damoradan, A.R. et al. Phase coexistence and electric-field control of toroidal order in oxide superlattices. \emph{Nature Materials} \textbf{16} 1003 (2017).
		
\bibitem{Hadjimichael18}
Hadjimichael, M., Zatterin, E., Fernandez-Pena, S., Leake, S.J., \&  Zubko, P. Domain Wall Orientations in Ferroelectric Superlattices Probed with Synchrotron X-Ray Diffraction \emph{Phys. Rev. Lett.} \textbf{120} 037602 (2018).
		
		
\bibitem{Callori12}
Callori, S.J., Gabel, J., Su, D., Sinsheimer, J., Fernandez-Serra, M.V., Dawber, M.. Ferroelectric PbTiO$_{3}$/SrRuO$_{3}$ superlattices with broken inversion symmetry \emph{Phys. Rev. Lett.} \textbf{109} 067601 (2012).

\bibitem{Sinsheimer12}
Sinsheimer, J, Callori, S.J., Bein, B., Benkara, Y., Daley, J., Coraor, J., Su, D., Stephens, P.W., \& Dawber, M. Engineering polarization rotation in a ferroelectric superlattice \emph{Phys. Rev. Lett.} \textbf{109} 167601 (2012).

\bibitem{Pertsev98}
Pertsev, N.A., Zembilgotov, A.G., Tagantsev, A.K. 	Effect of Mechanical Boundary Conditions on Phase Diagrams
of Epitaxial Ferroelectric Thin Films \emph{Phys. Rev. Lett.} \textbf{80} 1988 (1998).

\bibitem{Okatan09}
Okatan, M.B., Mantese, J.V, \& Alpay, S.P. Polarization coupling in ferroelectric multilayers \emph{Phys. Rev. B} \textbf{79} 174113 (2009).

\bibitem{Vlieg88}
Vlieg, E. et al., Surface X-ray scattering during Crystal Growth: Ge on Ge(111). \emph{Phys. Rev. Lett.} \textbf{61} 2241 (1988).
	

\bibitem{Levine89}
Levine, J.R. et al. Grazing-Incidence Small-Angle X-Ray Scattering: new tool for studying thin film growth. \emph{J. Appl. Crystallogr.} \textbf{22} 528-532 (1989).

 
\bibitem{Renaud03}
Renaud, G., Lazzari, R. et al. Real-time monitoring of growing nanoparticles. \emph{Science} \textbf{300} 1416 (2003).

 
\bibitem{Renaud09}
Renaud, G., Lazzari, R., Leroy, F. Probing surface and interface morphology with Grazing Incidence Small Angle X-Ray Scattering. \emph{Surf. Sci. Rep.} \textbf{64} 255-380 (2009).

 
\bibitem{Ozyadin05}
Ozaydin, G., Ozcan, A. S., Wang, Y., Ludwig, K. F., Zhou, H., Headrick, R. L. and Siddons, D. P. Real-time x-ray studies of Mo-seeded Si nanodot formation during ion bombardment. \emph{Appl. Phys. Lett.} \textbf{87} 163104 (2005).

\bibitem{Rainville15}
Rainville, M.G., Wagenbach, C., Ulbrandt, J.G., Narayanan, S., Sandy, A.R. Zhou, H., Headrick, R. L, and Ludwig, K.F. Co-GISAXS technique for investigating surface growth dynamics. \emph{Phys. Rev. B.} \textbf{92} 214102 (2015).

\bibitem{Ju19}
Ju, G., Xu, D., Highland, M.J., Thompson, C., Zhou, H., Eastman, J.A., Fuoss, P.H., Zapol, P., Kim, H. \& Stephenson, G.B. Coherent X-ray spectroscopy reveals the persistence of island arrangements during layer-by-layer growth. \emph{Nature Physics} https://doi.org/10.1038/s41567-019-0448-1 (2019).

\bibitem{Sinsheimer13}
Sinsheimer, J. et al. In-situ x-ray diffraction study of the growth of highly strained epitaxial BaTiO$_{3}$ thin films. \emph{Appl. Phys. Lett.} \textbf{103} 242904 (2013).


\bibitem{Murty02}
Ramana Murty, M. V. et al. In situ x-ray scattering study of PbTiO$_3$ chemical-vapor deposition. \emph{Appl. Phys. Lett.} \textbf{80} 1809 (2002).
	
	
\bibitem{Chinta12}
Chinta, P.V., Callori, S.J., Dawber, M., Ashrafi, A. \& Headrick. R.L. Transition from laminar to three-dimensional growth mode in pulsed laser deposited BiFeO$_{3}$ film on (001) SrTiO$_{3}$. \emph{Appl. Phys. Lett.} \textbf{101} 201602 (2012).

\bibitem{DRS05}
Dawber, M., Rabe, K.M. \& Scott, J.F. Physics of thin-film ferroelectric oxides. \emph{Rev. Mod. Phys.} \textbf{77} 1083 (2005).

\bibitem{Seidel09}
Seidel, J. et al. Conduction at domain walls in oxide multiferroics. \emph{Nat. Mater.} \textbf{8} 229–234 (2009).	

\bibitem{Guyonnet11}
Guyonnet, J., Gaponenko, I., Gariglio, S. \& Paruch, P. Conduction at Domain Walls in Insulating Pb(Zr$_{0.2}$Ti$_{0.8}$)O$_{3}$ Thin Films. \emph{Adv. Mater.} \textbf{23} 5377–5382 (2011).

\bibitem{Maksymovych11}
Maksymovych, P. et al. Dynamic Conductivity of Ferroelectric Domain Walls in BiFeO$_{3}$. \emph{Nano Lett.} \textbf{11} 1906–1912 (2011).

\bibitem{Farokhipoor11}
Farokhipoor, S. \& Noheda, B. Conduction through 71\textdegree\,  Domain Walls in BiFeO$_{3}$ Thin Films. \emph{Phys. Rev. Lett.} \textbf{107} 127601 (2011).


\bibitem{Catalan12}
Catalan, G., Seidel, J., Ramesh, R. \& Scott, J. F. Domain wall nanoelectronics. \emph{Rev. Mod. Phys.} \textbf{84} 119–156 (2012).

\bibitem{Sluka13}
Sluka, T., Tagantsev, A. K., Bednyakov, P. \& Setter, N. Free-electron gas at charged domain walls in insulating BaTiO$_{3}$. \emph{Nat. Commun.} \textbf{4} 1808 (2013).

\bibitem{Whyte15}
Whyte, J.R. \& Gregg, J.M. A diode for ferroelectric domain-wall motion. \emph{Nat. Commun.} \textbf{6} 7361 (2015).





	
\end{thebibliography}
\end{document}